\documentclass[english,prd,superscriptaddress,nofootinbib,preprintnumbers,twocolumn,showpacs]{revtex4}
\usepackage[latin1]{inputenc}
\usepackage{amsmath, amssymb, amsthm, amsfonts}
\usepackage{graphicx, bm, color, dcolumn, babel}

\makeatletter

\def\be{\begin{equation}}
\def\ee{\end{equation}}
\def\ba{\begin{eqnarray}}
\def\ea{\end{eqnarray}}

\begin{document}

\title{Local and non-local measures of acceleration in cosmology}

\author{Philip Bull}
\email{Phil.Bull@astro.ox.ac.uk}
\affiliation{Department of Astrophysics, University of Oxford, UK.}

\author{Timothy Clifton}
\email{Tim.Clifton@astro.ox.ac.uk}
\affiliation{Department of Astrophysics, University of Oxford, UK.}
\affiliation{School of Physics and Astronomy,
 Queen Mary University of London, UK.}

%----------------------------------------------
\begin{abstract}
%----------------------------------------------

Current cosmological observations, when interpreted within the framework of a 
homogeneous and isotropic Friedmann-Lema\^{i}tre-Robertson-Walker
(FLRW) model, strongly suggest that the Universe is entering a period of
accelerating expansion.  This is often taken to mean that the
expansion of space itself is accelerating.  In a general spacetime,
however, this is not necessarily true.  We attempt to 
clarify this point by considering a handful of local and non-local measures 
of acceleration in a variety of inhomogeneous cosmological models. Each of the 
chosen measures corresponds to a theoretical or observational procedure that 
has previously been used to study acceleration in cosmology, and all
measures reduce to the same quantity in the limit of exact spatial
homogeneity and isotropy.  In statistically homogeneous and isotropic
spacetimes, we find that the acceleration inferred from observations of
the distance-redshift relation is closely related
to the acceleration of the spatially averaged universe, but does not necessarily bear
any resemblance to the average of the local acceleration of spacetime itself. 
For inhomogeneous spacetimes that do not display
statistical homogeneity and isotropy, however, we find little correlation between
acceleration inferred from observations and the acceleration of the
averaged spacetime.  This shows that observations made in an
inhomogeneous universe can imply acceleration without the existence of dark energy.

%----------------------------------------------
\end{abstract}
%----------------------------------------------
%\date{\today}
\maketitle

%----------------------------------------------
\section{Introduction}\label{sect-intro}
%----------------------------------------------

The question of if, and why, the expansion of the Universe is accelerating is 
one of the foremost problems in fundamental physics today. 
Observations of distant Type Ia supernovae are well-fit by homogeneous and
isotropic Friedmann-Lema\^{i}tre-Robertson-Walker (FLRW) solutions of general 
relativity, but only if a substantial fraction of the energy density in the 
Universe is in the form of a cosmological constant. For this constant 
to take the value required by observations, however, requires an extraordinary 
degree of fine-tuning. Many authors have attempted to remedy this situation 
by proposing modifications to gravity, the existence of negative-pressure 
components of the cosmological fluid, new scalar fields, and a host of other 
exotic mechanisms. Others have sought to explain it by modifying the 
basic assumptions of cosmology itself, such as the Cosmological Principle.

Our aim here is to clarify what is meant by acceleration in an
inhomogeneous universe, and to study how different types of acceleration can 
arise within the context of relativistic cosmology. We do not
consider the existence of any strange new matter fields or
modifications to Einstein's equations, but we do
allow for the Universe to be strongly inhomogeneous below a certain
scale.  It is found that observations made over
large distances (e.g. using supernova and CMB observations) are best
modeled using non-local averages of geometric quantities.  The
acceleration that one can infer from these large-scale probes of
cosmology is not necessarily similar to any locally defined measures
of acceleration, and does not necessarily imply that the expansion of
space itself is accelerating at any point.

Following Hirata \& Seljak \cite{Hirata} and Clarkson \& Umeh \cite{CU}, we 
consider multiple different measures of acceleration:
\begin{enumerate}
 \item[(a).] The local acceleration of nearby observers in a small volume of space,
   as governed by the Raychaudhuri equation;
 \item[(b).] The acceleration inferred by fitting an FLRW 
 distance-redshift relation to Hubble diagrams constructed from
 observations made over large distances;
 \item[(c).] The acceleration inferred by constructing a Hubble diagram 
 from local observations (i.e. at $z\simeq 0$), as described by Kristian
 and Sachs \cite{KS};
 \item[(d).] The average acceleration of a large volume of space, as calculated
   using Buchert's scalar averaging procedure \cite{Buchert1}.
\end{enumerate}
For an exactly homogeneous and isotropic FLRW model, all of these measures
are identical. In the general case,  however, they can be quite
different, to the point that some of them can  show strong
acceleration, while others show deceleration.

Measures (a) and (c) are purely local, and depend 
only on the curvature of spacetime within the neighbourhood of a point (in the limit of
geometric optics).   Measures (b) and (d), however, are both
inherently non-local in nature.
Measure (a) describes the actual, dynamical acceleration of 
a volume of space, while measure (b) corresponds to what is measured in, for example,
supernova surveys.  Measure (c) is a local measure of acceleration,
based on inferences made using the luminosity distance-redshift
relation within a very small region of spacetime.
Measure (d) corresponds to the theoretical procedure of fitting a smooth 
effective model to the real, inhomogeneous Universe 
(the question of how best to do this is still very much an open one, but the simplest
and best-known procedure is the scalar averaging formalism
developed by Buchert \cite{BuchertBR1,BuchertBR2}).
We will discuss all of these measures in detail in what follows.

Constructing relativistic models of the real, inhomogeneous Universe is 
difficult, but a number of attempts have met partial success in addressing 
the problem. Realistic density fields can be described using perturbations to 
exact FLRW solutions. However, at least to quadratic order in
perturbations, the effect of inhomogeneities on the overall behaviour
of the spacetime often turns out to be small, with the average evolution and 
optical properties of the model remaining close to the unperturbed background 
values \cite{IshibashiWald, GreenWald, Kolb2, Paranjape3, Kasai, RasanenNewtonian, Flanagan, Hirata, Geshnizjani, Weinberg, Bonvin, Baumann}. 
There have been suggestions that effects at higher orders in perturbations 
could be important \cite{CU,RasanenNewtonian}, and recent studies of 
non-linear collapse which make use of a gradient expansion also find potentially 
significant effects \cite{Grad1, Grad2}. These claims require 
further analysis if they are to be confirmed, however. The main drawback of the 
perturbative approach is that the solutions obtained are not exact, and so 
it is unclear if we are neglecting or incorrectly estimating aspects of the 
fully-relativistic behaviour of the spacetime.

An alternative approach involves studying exact, inhomogeneous
solutions to Einstein's equations. As the geometry of spacetime is known from the
outset, the non-linear evolution of space, and light rays within 
it, can be calculated in a fully rigorous manner, without recourse to 
perturbative analyses.
This approach tends to require a high degree of symmetry, however, placing 
strong restrictions on what it is possible to model \cite{EllisSC}.
For example, backreaction effects in Swiss Cheese models (constructed by joining together
spherically-symmetric inhomogeneous regions with a FLRW solution) are known to be small 
\cite{Mattsson}.  This is confirmed by ray tracing studies through a variety 
of similar exact solutions 
\cite{Mattsson, Meures, DiDio, Szybka, Brouzakis, Vanderveld, BolejkoSC1, CliftonZuntz}. 
Additionally, inhomogeneous models with general fluid content 
are difficult to work with \cite{BolejkoLasky, ClarksonRegis, Marra}, and so 
most modeling attempts make do with a dust-only stress-energy tensor. This 
is itself problematic, as high-density regions tend to rapidly collapse
in inhomogeneous pressureless fluids, forming singularities 
and resulting in unrealistic, pathological behaviour.

Finally, more heuristic analyses have been attempted
\cite{LindquistWheeler, CliftonFerreira, Mattsson, Rasanen2008, Kantowski, RasanenBR, Rasanen2, Rasanen3}. These often employ disjoint 
regions of different exact solutions. The dynamical evolution within each 
region is therefore well defined, but the ensemble as a whole does not satisfy 
Einstein's equations at the boundaries between regions.
This approach relaxes the restrictive symmetry requirements that are
necessary when using exact solutions, but 
introduces its own ambiguities. Indeed, the chosen 
boundary conditions can sometimes dominate the behaviour of the model 
\cite{CliftonFerreira}, and it is not clear whether the resulting spacetime 
approximates any actual solution of Einstein's equations or not.

In this paper we consider the latter two of these approaches, as the situation of
small perturbations around an exact FLRW geometry is well-studied
already \cite{Kolb1, Kolb2, Paranjape2, Brown, Li, Bonvin}.
The format of the paper is as follows. In Section \ref{sect-measures}, we 
discuss in detail the different measures of acceleration summarised above, 
and how they can be 
calculated in a general spacetime. The conditions for these measures to show 
acceleration, and their relation to observable quantities (if any), is 
discussed. Section \ref{sect-models} sets out three different inhomogeneous 
models: the spherical collapse model, constructed from disjoint FLRW regions; 
the Kasner-EdS model, an exact solution with alternating expanding 
vacuum and collapsing dust regions along a line of sight 
\cite{Dyer1, Lake, Landry, Dyer, Hellaby}; and the 
Lema\^{i}tre-Tolman-Bondi model, an exact, spherically-symmetric 
dust solution \cite{LTB1, LTB2, LTB3}.  The
volume of space in all of these models is locally 
decelerating everywhere, and yet we find that observations made within
them can still exhibit acceleration.

In Sections \ref{sect-results-sc}, \ref{sect-results-kasner}, and 
\ref{sect-results-ltb}, we present our results for each of the models. We find that the acceleration 
inferred by observers from the Hubble diagram, constructed over large
distances, is most closely correlated 
with the dynamical behaviour of the {\it averaged} spacetime, and not with its 
local acceleration. The local acceleration, however, is well modeled
by Hubble diagrams that are constructed locally, using the
Kristian-Sachs formalism.  This result suggests that we should use
non-local averages of the geometry to interpret observations made over
large distances, and local geometry to interpret observations made locally.
Interpreting observations within the wrong framework can lead to
incorrect inferences about what is causing the acceleration. This appears to 
us to provide some insight into the question of whether the apparent 
acceleration we observe is necessarily caused by the presence of a non-zero 
cosmological constant, or whether it could be caused by some backreaction 
effect within the inhomogeneous spacetime. Finally, in 
Section \ref{sect-discussion} we discuss our results in the context of 
previous claims involving backreaction and the nature of the 
apparent acceleration of the Universe.

%----------------------------------------------
\section{Measures of acceleration}\label{sect-measures}
%----------------------------------------------

In this section we discuss the four measures of acceleration 
that were listed in Section \ref{sect-intro}. Each 
measure has associated with it a {\it deceleration parameter}. For each 
definition, we set out its theoretical basis and physical interpretation, list 
the conditions necessary for acceleration to occur, and describe the relation 
of the deceleration parameter associated with this measure to observable quantities. The measures are 
summarised in Table \ref{tbl-summary}, at the end of the section.

\subsection*{(a) Local volume acceleration} \label{sect-def-qloc}

A congruence of time-like geodesics, $u^a$, describing the world-lines 
of a set of observers comoving within a cosmological fluid can be
decomposed such that~\cite{EllisVarenna}
\be
u_{a;b} = \frac{1}{3} \Theta h_{ab} +\sigma_{ab} +\omega_{ab},
\ee
where subscript $;$ denotes a covariant derivative, and $h_{ab}=g_{ab}+u_a u_b$
is the projection tensor.
The kinematic quantities in this equation are the expansion scalar,
$\Theta$, the shear, $\sigma_{ab}$, and the vorticity, $\omega_{ab}$,
which correspond to the trace, symmetric trace-free, and antisymmetric
parts of $u_{a;b}$, respectively.  This decomposition 
is a fully covariant procedure, valid for any spacetime. 

For irrotational flows ($\omega_{ab}=0$), the fluid flow becomes
hypersurface orthogonal, with the projection tensor becoming
the induced metric of the orthogonal 3-spaces.  The field equations 
then give (with $8\pi G = 1$)
\be
\dot{\Theta} =-\frac{1}{3}\Theta^2 - \frac{1}{2}(\rho + 3p) +
\Lambda - 2\sigma^2,  \label{eqn-raychoudhuri}
\ee
where the over-dot denotes a derivative with respect to proper time along $u^a$, and where
$\sigma^2=\frac{1}{2}\sigma_{ab}\sigma^{ab}$, and $\rho$, $p$
and $\Lambda$ are the energy density, pressure and cosmological
constant, respectively.  The Gauss embedding equation for these
hypersurfaces is
\be
\frac{1}{3}\Theta^2 = \rho + \Lambda + \sigma^2 - \frac{1}{2}
     {^{(3)}\mathcal{R}}, \label{eqn-friedmann}
\ee
where ${^{(3)}\mathcal{R}}$ is the Ricci scalar constructed from $h_{ab}$.

By analogy with the deceleration parameter in an FLRW spacetime, one
can then define a local volume deceleration parameter,
\be
q_{\Theta} = -1 -3 \frac{\dot{\Theta}}{\Theta^2} 
 = \frac{3}{\Theta^2} \left [ \frac{1}{2}(\rho + 3p) - \Lambda + 2
  \sigma^2 \right ],
\label{eqn-qloc}
\ee
corresponding to the monopole of the deceleration of the expansion rate of a set of
neighbouring particles following $u^a$. This measure is said to
be accelerating when $q_{\Theta} < 0$, which
from Eq. (\ref{eqn-qloc}) can be seen to occur if and only if
$p<-\frac{1}{3}\rho$ or $\Lambda > 0$. 

Now, $q_{\Theta}$ is a local measure of acceleration, defined only in the
neighbourhood of a point in spacetime. 
Determining $q_\Theta$ from observations therefore requires $\Theta$ (and its first
derivative) to be determined using observations within the
neighbourhood of a single point in spacetime only.  These observations
need to be of the rate of change of proper distance between particles in their own
rest-frame, which is not necessarily the same as the angular diameter
distance or luminosity distance (for these see measure (c), below).
The Universe is extremely inhomogeneous on small scales, and 
so this measure of acceleration is likely to display considerable spatial 
variation. A local measurement of $q_\Theta$ need not, then, be representative 
of the mean local volume acceleration.

\subsection*{\qquad (b) Observed acceleration \newline (as inferred from the Hubble diagram)} \label{sect-def-q0}

To date, the strongest observational evidence for an accelerating
universe comes from the distance-redshift relation, which is measured
using ``standardisable candles'' (such as Type Ia supernovae), as well
as the CMB.  By fitting this data to the relations derived from the FLRW 
solutions of Einstein's equations, one finds that models with 
$\Omega_\Lambda > 0$ are strongly favoured, while those with 
$\Omega_\Lambda = 0$ are ruled out to a high degree of confidence. Under the 
assumption that spacetime is well-described by FLRW, this constitutes 
strong evidence for accelerating expansion.

A key step in this procedure is the fitting of the FLRW distance-redshift 
relation to data. Performing a series expansion in the angular diameter 
distance about $z=0$ gives
\be \label{eqn-dist-expansion}
d_A(z) = \frac{c z}{H_0} \left ( 1 - \frac{1}{2}(3 + q_0)z + \mathcal{O}(z^2) \right ),
\ee
where $H_0$ is the inferred Hubble rate at $z=0$, and $q_0$ is the inferred deceleration parameter.  The value of
$q_0$ can be found in a purely observational manner by taking derivatives of the fitted 
distance-redshift relation \cite{Visser},
\be\label{eqn-q0}
q_0 = -\left . \frac{d^{\prime\prime}_A}{d_A^\prime} \right |_0 - 3,
\ee
where primes denote derivatives with respect to redshift, and where subscript 
$0$ denotes that a quantity is evaluated at the observer.  Note that
this measure of deceleration is inferred from observations over large
distances, and so assigns information to a point in spacetime based on
information obtained from the entire extended region over which observations have
been made.  This is a highly non-local process.

Eq. (\ref{eqn-q0}) is derived from the distance-redshift relation in
FLRW geometry, but one should note that the existence of such a
geometry is not required in order to infer $q_0$ from
observations.  What we are doing here should instead be considered
simply as a fitting procedure. 
Eq. (\ref{eqn-q0}) is then an observable in {\it any} spacetime (after
averaging over the celestial sphere), and so anyone can 
measure $q_0$ if they are willing to interpret their observations within the 
framework of an FLRW model.
We call the acceleration inferred by this fitting of FLRW relations to the 
monopole of the distance-redshift relation the {\it observed acceleration}.

The distance-redshift relation can be calculated in a general spacetime by 
solving the Sachs optical equations for bundles of null geodesics.
With vanishing vorticity, these are
\ba
\frac{d\theta}{d\lambda} + \theta^2 + |\hat{\sigma}|^2 &=& -\frac{1}{2}
R_{ab} k^a k^b \label{sachs-theta} \label{eqn-sachs-theta} \\
\frac{d\hat{\sigma}}{d\lambda} + 2 \theta \hat{\sigma} &=& C_{abcd}
(t^*)^a k^b (t^*)^c k^d\label{sachs-sigma},
\ea
where $\lambda$ is an affine parameter along the bundle, $R_{ab}$ and 
$C_{abcd}$ are the Ricci and Weyl tensors of the spacetime, $\theta$ and 
$\hat{\sigma}$ are the expansion and (complex) shear scalars of the null 
geodesics, $k^a$ is a tangent 
vector to the null curves, and $t^a$ are (complex) vectors spanning a two-dimensional 
screen space orthogonal to $k^a$. The expansion scalar is related to 
the angular diameter distance measured along the bundle by
$\theta = d (\ln (d_A))/d\lambda$.
Substituting this into Eq. (\ref{eqn-sachs-theta}) yields
\be
\label{eqn-sachs-theta2}
\frac{d^2 (d_A)}{d\lambda^2} = -d_A \left ( |\hat{\sigma}|^2 + \frac{1}{2}R_{ab}k^a k^b\right ).
\ee
The luminosity distance is related to the angular diameter distance by
the reciprocity theorem \cite{Etherington}, which gives $d_L=(1+z)^2 d_A$.

The affine distance-redshift relation for a general spacetime is given by 
\cite{ClarksonMaartens}
\be \label{eqn-redshift}
\frac{dz}{d\lambda} = -(1+z)^2 H_{||}(z),
\ee
where $H_{||}=\frac{1}{3} \Theta+ \sigma_{ab} e^a e^b$ is the expansion
rate along the line of sight, $e^a$. Equations 
(\ref{sachs-sigma})--(\ref{eqn-redshift}) can be solved to give 
the angular diameter distance-redshift relation, $d_A(z)$, in any given direction, 
at any given point in spacetime. This procedure can be repeated for every direction 
on the sky, and an FLRW model can be fitted to the monopole of the resulting angular 
distribution. The best-fit model can then be used to find $q_0$. We
denote the result of this procedure in a general spacetime as  $q_\mathrm{obs}$.

We consider this definition of acceleration to correspond most closely to the one used by 
observers. It is a non-local measure, since it depends on solutions to the 
Sachs equations, which describe bundles of null geodesic curves that
extend through the 
spacetime. In effect, the measurement of $q_\mathrm{obs}$ depends on finding 
the entire past null cone of an observer out to some $z$, and 
fitting some distance-redshift relation to it. The conditions for a spacetime 
to have $q_\mathrm{obs} < 0$ are 
therefore complicated, in general. As we will see below, it is possible 
to find spacetimes that are quite different from simple $\Lambda$-dominated FLRW models 
that nevertheless have $q_\mathrm{obs} < 0$.

\subsection*{\qquad (c) Acceleration from local observations \newline (using the
  Kristian-Sachs formalism)}

The measure of acceleration we just described has the disadvantage of requiring solutions to the 
Sachs equations to be found, as a function of redshift and angle on the 
observer's sky. This can be a difficult task in general,
as it requires detailed knowledge of the geometry of spacetime. Instead, 
one can use the Kristian-Sachs formalism \cite{KS} to obtain a fully general 
and covariant series expansion of the distance-redshift relation about
an observer without needing 
to consider solutions to the geodesic equations at all. Furthermore, the 
expansion can be decomposed directly into covariant spherical harmonics about 
the observer, allowing the monopole term to be calculated straight away \cite{CU}.

This procedure has been spelled out in detail by Clarkson and Umeh \cite{CU}. The generalised form 
of Eq. (\ref{eqn-dist-expansion}) is
\be \label{eqn-ks-da}
d_A = \frac{z}{[K^a K^b \nabla_a u_b]_0} \left ( 1 - \left [\frac{K^a
    K^b K^c \nabla_a \nabla_b u_c }{2(K^d K^e \nabla_d u_e)^2} \right
]_0 z + \mathcal{O}(z^2) \right ), 
\ee
where the past-pointing null direction can be written in terms of the tangent 
vector to a comoving observer's world-line, $u^a$, and a direction on their 
sky, $e^a$, as
\be
K^a = \frac{k^a}{[u_b k^b]_0} = -u^a + e^a.
\ee
The subscript $0$ again denotes evaluation at the observer's location. 
The terms in Eq. (\ref{eqn-ks-da}) can be expanded using a covariant 
decomposition in spherical harmonics. In order to facilitate this expansion, it is useful
to invert Eq. (\ref{eqn-ks-da}) to give
\be \label{eqn-ks-z}
z = [K^a K^b \nabla_a u_b]_0 d_A + \frac{1}{2}[K^a K^b K^c \nabla_a \nabla_b u_c]_0 d_A^2 + \mathcal{O}(d_A^3).
\ee
All of the terms that we wish to expand are now in the numerator. 
Comparing the monopoles of the coefficients in Eq. (\ref{eqn-ks-z}) with the
corresponding FLRW relation (with $8\pi G = 1$) then gives
\be \label{eqn-qks}
q_{\mathrm{KS}} = \frac{3}{\Theta^2} \left [ \frac{1}{2}(\rho + 3p) - \Lambda + 6\sigma^2  \right ]_0,
\ee
where we have used $H_0 = \frac{1}{3}\Theta$, which corresponds to the
monopole of this term, rather than its full spherical harmonic 
expansion. This corresponds most closely with the way that $H_0$ is typically 
used in observational studies; the monopole of $H_0$ tends to be determined 
separately from other quantities.
From Eq. (\ref{eqn-qks}), it can be seen that $q_{\mathrm{KS}} \ge 0$
unless $p<-\frac{1}{3} \rho$ or $\Lambda > 0$.  That is, 
acceleration of this measure can only occur if there is a cosmological
constant, or if an exotic fluid with negative pressure is present.

There is clearly some similarity between the deceleration parameter of the local
volume, $q_{\Theta}$, and that which is obtained from the local
distance-redshift relation, $q_{\mathrm{KS}}$. Both 
measures of acceleration are local (depending only on quantities defined 
within the neighbourhood of the observer), and both are given by
expressions that differ only by a term involving the shear scalar.
This does not, however, mean that these two measures of acceleration
are the same.  They correspond to different physical quantities.

The relation between the Kristian-Sachs and observational acceleration 
measures is less clear. This has to do with the non-locality of the latter; if 
the Hubble diagram could be measured precisely at $z=0$ (i.e. in 
the limit of geodesics of zero affine length), we would find 
$q_{\mathrm{obs}} = q_{\mathrm{KS}}$. 
But, because real observations necessarily cover a range of redshifts, the 
process of fitting a curve to the data and extrapolating that back to $z=0$ 
means that in general we will have $q_{\mathrm{obs}} \neq
q_{\mathrm{KS}}$.  This phenomenon has been investigated 
using real data in \cite{CFL}. As we shall see in what follows, even the 
signs of $q_{\mathrm{obs}}$ and $q_{\mathrm{KS}}$ can be different.
That is, a locally-decelerating 
spacetime can still have a Hubble diagram that implies acceleration.

\subsection*{\qquad (d) Acceleration of the average \newline (using Buchert's formalism)} \label{sect-def-qd}

One method of constructing a 
homogeneous and isotropic `effective' model within which observations
can be interpreted involves taking 
averages of geometrical quantities over space-like hypersurfaces. The hope is then
that the behaviour of the averaged model will capture some aspects of the real spacetime, 
both in terms of its dynamics, and the observational
quantities that are calculated within it. Many such averaging procedures exist 
\cite{Hoogen}, but here we will concentrate on the Buchert scalar 
averaging formalism \cite{Buchert1}. This is the most widely
used formalism in the literature.

Buchert's method proceeds as follows. First of all, the spacetime is
filled with a congruence of (irrotational) curves.  It is then
foliated with a set of space-like hypersurfaces orthogonal to these curves.
The proper 3-volume of a domain, $\mathcal{D}$, on 
a given hypersurfaces is
\be
V_\mathcal{D} = \int_\mathcal{D} \sqrt{-h}~d^3x,
\ee
where $h_{ij}$ is the induced metric on the hypersurface, and $h =
\mathrm{det}~h_{ij}$. In general, the induced metric will be a 
function of time, and so the volume is time-dependent as well. The 
proper volume-weighted average of a scalar quantity, $S$, over a
spatial domain, $\mathcal{D}$, can then be written as
\be \label{eqn-buchert-avg}
\langle S \rangle = V^{-1}_\mathcal{D} \int_\mathcal{D} S(\vec{x}, t) \sqrt{-h}~d^3x.
\ee
Spatial averaging and time evolution do not, in general, commute.
They instead obey the commutation relation
\be \label{eqn-avg-commutation}
\partial_t \langle S \rangle - \langle \partial_t S \rangle = \langle
\Theta S \rangle - \langle \Theta \rangle \langle S \rangle. 
\ee
An `effective' homogeneous model can be constructed by averaging over domain 
sizes greater than the statistical homogeneity scale of the underlying 
inhomogeneous spacetime. An effective scale factor for the resulting model can
then be defined as
\be \label{eqn-avg-a}
a_\mathcal{D}(t) = \left ( \frac{V_\mathcal{D}(t)}{V_\mathcal{D}(t_0)} \right )^\frac{1}{3},
\ee
where $t_0$ is some fiducial time.  For geodesic curves, and
pressure-free matter, we can use this formalism to construct analogues to the 
Friedmann and Raychaudhuri equations,
\ba
3 H^2_\mathcal{D} 
 &=& 8\pi G \langle\rho\rangle + \Lambda - \frac{1}{2} \left (Q_\mathcal{D} + \langle ^{(3)}\mathcal{R} \rangle \right ) \nonumber\\
3 \frac{\ddot{a}_\mathcal{D}}{a_\mathcal{D}} &=& -8 \pi G \langle \rho \rangle + \Lambda + Q_\mathcal{D} \label{eqn-buchert-raych},
\ea
where $H_{\mathcal{D}}=\dot{a}_{\mathcal{D}}/a_{\mathcal{D}}$.
Over-dots denote partial differentiation with respect to $t$, the
proper time along curves orthogonal to the 3-space.
The {\it kinematical backreaction scalar} is defined to be
\be \label{eqn-Q}
Q_\mathcal{D} = \frac{2}{3}\left ( \langle \Theta^2 \rangle - \langle \Theta \rangle^2\right ) - 2\langle \sigma^2 \rangle.
\ee
A deceleration parameter for the averaged hypersurfaces can then be defined, by 
analogy with FLRW cosmological models, as
\be \label{eqn-qd}
q_\mathcal{D} = -\frac{1}{H^2_\mathcal{D}}\frac{\ddot{a}_\mathcal{D}}{a_\mathcal{D}}.
\ee
An effective distance-redshift relation can also be found for the averaged 
model by assuming that light rays follow null geodesics of the
averaged spacetime, and that geodesic observers are comoving in the
average geometry.

In general, the deceleration parameter $q_\mathcal{D}$ is non-local, and 
not directly observable, as it depends on averages over extended space-like hypersurfaces. From 
Eq. (\ref{eqn-buchert-raych}), one can see that if the averaged spatial Ricci 
curvature, $\langle ^{(3)}\mathcal{R}\rangle$, and backreaction, $Q_\mathcal{D}$,
behave in a certain way, then it is possible to have $q_\mathcal{D}<0$
without having $\Lambda>0$ or $\langle \rho \rangle <0$. In particular, spacetimes 
consisting of collapsing regions in an expanding background can exhibit this 
behaviour, which has led some to claim that the apparent cosmic acceleration 
inferred from supernova observations could instead be explained as a 
consequence of the variance of the inhomogeneous expansion rate that enters 
into the definition of $Q_\mathcal{D}$ \cite{BuchertBR1, BuchertBR2, Rasanen1, RasanenBR, Barausse, Kolb1}. We will examine this claim in detail later on, 
but for now will just note that it is possible for the Buchert averaged model 
to accelerate, even if $q_\Theta > 0$ everywhere.

\begin{table}[h]
 \begin{tabular}{|l|c|l|l|}
 \hline
 {\bf Measure} & {\bf Local} & {\bf Support} & {\bf Observable} \\
 \hline
 Local volume, $q_\Theta$ & Yes & Spacetime point & In principle \\
 Observational, $q_\mathrm{obs}$ & No & Null geodesic & Yes \\
 Kristian-Sachs, $q_{\mathrm{KS}}$ & Yes & Spacetime point & In principle \\
 Buchert average, $q_\mathcal{D}$ & No & Spatial domain & No \\
 \hline
 \end{tabular}
 \caption{Summary of the different measures of acceleration 
 defined in Section \ref{sect-measures}.}
 \label{tbl-summary}
\end{table}

Quantities of particular interest are the deceleration parameters that observers
in a given region of spacetime 
should {\it expect} to infer, rather than the actual values of these
parameters at single points (which 
may not be representative). To this end, we also use Eq. 
(\ref{eqn-buchert-avg}) to average our various measures of acceleration. For 
example, $\langle q_{\mathrm{KS}} \rangle$ will be taken to correspond to the 
value of the Kristian-Sachs deceleration parameter that a typical observer would 
expect to measure in a given region. Unless specified otherwise, the averaging 
domain is taken to be larger than the homogeneity scale of the model\footnote{
It should be noted that this averaging scheme is weighted by 
proper volume, and so the density of observers is implicitly assumed to be 
weighted in a similar manner. Of course, this need not be the case, and 
other ways of distributing observers throughout space could be considered. 
This would affect what a `typical' observer should expect to see. We do not 
consider this question further here.}.

\begin{figure*}
 \centerline{
  \includegraphics[width=14cm]{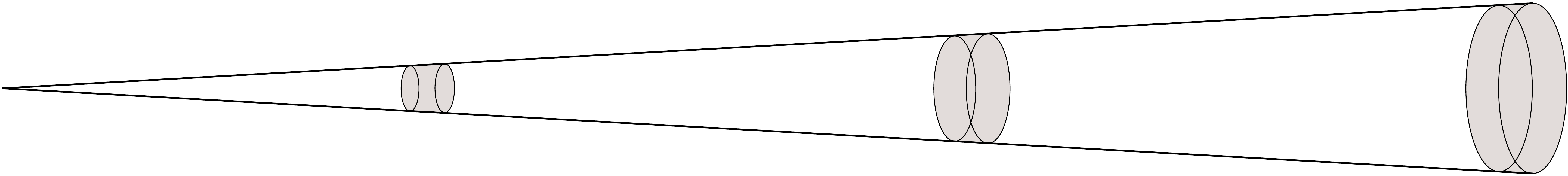} }
 \caption{Schematic representation of a line-of-sight through a universe
   with alternating expanding vacuum regions (clear), and
 collapsing dust regions (grey).}
 \label{fig-beam}
\end{figure*}

%----------------------------------------------
\section{Inhomogeneous cosmological models}\label{sect-models}
%----------------------------------------------

We now consider a number of different relativistic models, and how the 
measures of acceleration defined in Section \ref{sect-measures} can be calculated 
in each of them. We will primarily be interested in cases with vanishing 
cosmological constant, as a non-zero $\Lambda$ generically results in accelerating 
expansion in all of the measures discussed above. The models we
consider are chosen as illustrative examples, as they are capable of
showing acceleration in some measures, while displaying
deceleration in others.

\subsection{Spherical collapse model} \label{sect-model-sc}

In this section we consider the spherical collapse model, which consists of 
an ensemble of disjoint FLRW regions \cite{GunnGott, Padmanabhan, Fosalba, Abramo}. 
It is not an exact solution to Einstein's 
equations (distinct FLRW models cannot satisfy the appropriate junction conditions on their 
shared boundary), but provides a useful toy model of collapsing structures in 
an otherwise expanding universe. By virtue of each individual region being
homogeneous, spatial averages take a particularly simple form in 
spherical collapse 
models. A scalar quantity $S$ may be averaged simply by taking the sum of its 
values in the different regions weighted by the proper 3-volume of
each region, $V$, at a 
given time \cite{Rasanen2008, IshibashiWald},
\be
\langle S \rangle = \frac{\sum_i S_i V_i  }{ \sum_i V_i  }.
\ee
We wish to model the collapse of structures in an expanding background, as 
this configuration is known to lead to acceleration of the Buchert averaged 
spacetime, $q_\mathcal{D} < 0$ \cite{Rasanen2008, IshibashiWald}. To this end, 
we select alternating regions of collapsing, spatially closed and
dust-dominated regions, and expanding, spatially open vacuum regions,
as illustrated in Fig. \ref{fig-beam}.
We model the regions using FLRW geometry (with $\Lambda=0$), and
refer to them as Regions I and II, respectively. The collapsing regions are 
chosen so that they have a comoving size of order $10\%$ of the vacuum
regions at the present time.  As we look further back in time, the
collapsing regions increasingly dominate the proper spatial volume. 

In what follows we will use the following form for the FLRW metric,
\be
\label{FLRWds}
ds^2 = dt^2 - \frac{a^2 \left ( dX^2 + dY^2 + dZ^2\right )}{[1 + \frac{k}{4}(X^2 + Y^2 + Z^2)]^2},
\ee
and take the individual domains to have depth $X_\mathcal{D}$. 
The proper volume of each domain is
\be
V_i = \int_{\mathcal{D}_i} \frac{a_i^3 dX dY dZ}{[1 + (k_i/4)(X^2 + Y^2 + Z^2)]^3},
\ee
and the comoving depth of a domain is
\be
\chi_i = \int_0^{X_{\mathcal{D}_i}} \frac{dX}{1 + (k_i/4)(X^2 + Y^2 + Z^2)}.
\ee
The effective scale factor, defined in Eq. (\ref{eqn-avg-a}), is 
\be
a_\mathcal{D}(t) = \left ( \frac{\sum_i V_i(t)}{\sum_i V_i(t_0)} \right )^\frac{1}{3},
\ee
and the Buchert average deceleration parameter $q_\mathcal{D}$ is calculated 
according to Eq. (\ref{eqn-qd}).

The distance-redshift relation is required to find $q_\mathrm{obs}$. It can be 
found by solving the Sachs optical equations for a ray passing through a 
number of different FLRW regions. All that is required is to ensure 
continuity of the affine parameter, $\lambda$, the redshift, $z$, the
angular diameter distance, $d_A$, and the expansion scalar, $\theta$,
at the boundary, as a ray leaves one 
region and enters another. These conditions impose no restrictions on the 
individual FLRW regions.  Their size, matter content, and expansion rate can 
be chosen separately. We choose the cosmic time in individual regions such 
that it is continuous on the boundary between them.

The Sachs equations in each region are given by
\be\label{eqn-flrw-sachs}
\frac{d^2 (d_A)}{d\lambda^2} = -4\pi G \rho_i~ (1+z)^2 d_A,
\ee
with 4-vector tangent to the null curves given by
\be 
k^a= \left( 1 + z , \pm \frac{1+z}{a_i}\sqrt{1 + \frac{1}{4}k_i X^2}
,0,0 \right).
\ee
The redshift as a function of affine distance is then
\be
\label{zSC}
\frac{dz}{d\lambda} = -H_i (1+z)^2.
\ee
We have not used $1+z = a^{-1}$, as the total redshift along a 
ray, $z(\lambda)$, is not the same as the redshift that one would obtain by 
ray-tracing through a single FLRW region.

We propagate the light rays through each region in turn, until the
edge of the region is reached at some coordinate distance $X_i$ (which
we call the `length' of the region). This length is a function of time
(evaluated at the instant the ray enters a region), chosen so that the ray 
travels a proper distance in each region proportional to the fraction of the 
proper volume taken up by that region on a surface of constant $t$, i.e.
\be\label{eqn-sc-rayfraction}
\frac{a_I \chi_I}{a_{II} \chi_{II}} = \frac{V_I}{V_{II}}.
\ee
This choice takes into account not only the different expansion rates
of the two different regions along the line of sight, but also the
fact that as we go forwards in time the number density of dust regions
should be expected to decrease.  For simplicity, we fix the (comoving)
length of the collapsing region, $\chi_I$, and then solve
Eq. (\ref{eqn-sc-rayfraction}) for $\chi_{II}$ on entering each new region.

The redshift through the model is given by integrating Eq. (\ref{zSC}).
There is a blueshift as the photons pass through collapsing
regions, and a redshift as they pass through 
expanding regions. The distance increases despite the blueshift, and the 
distance-redshift relation becomes multi-valued. This leads to it taking on a 
jagged appearance, although the distance remains smooth 
as a function of affine parameter. The observational deceleration
parameter can be 
calculated by fitting an FLRW model with dust, curvature, and $\Lambda$ to the 
jagged distance-redshift curve, using a simple least-squares
procedure. The FLRW deceleration parameter, defined by
Eq. (\ref{eqn-q0}), can then be calculated
to give $q_\mathrm{obs}$.

The local volume deceleration parameter in each FLRW region is given by 
Eq. (\ref{eqn-qloc}). In our chosen spherical collapse model, 
$\Omega_\Lambda=0$ in both regions, and $\Omega_m = 0$ in the vacuum region, so 
the average of the local deceleration parameter reduces to
\be
\langle q_\Theta \rangle = \left . \frac{1}{2}\frac{\Omega_{m,I} V_{I}}{V_I + V_{II}} \right |_0,
\ee
which is positive definite (i.e. decelerating everywhere). The equality 
$q_\Theta = q_{\mathrm{KS}}$ holds, since each region is FLRW, and therefore 
has vanishing shear.

\subsection{Kasner-EdS model}\label{sect-model-kasner}

In the previous section we considered the spherical collapse model, 
which consists of disjoint regions of different FLRW
spacetimes.  This model is simple to work with, but only forms an approximate
solution of Einstein's equations.  In this section we consider an
exact solution that is inhomogeneous along the line of sight, with
alternating regions of collapsing dust and expanding vacuum, as
illustrated in Figure \ref{fig-beam}.  We take the two space-like
directions orthogonal to the line of sight to span a plane symmetric
subspace, and enforce statistical homogeneity along the line of sight
only.

In order to prevent the rapid formation of singularities that often
occurs when dust is allowed to collapse in general relativistic models,
we will take the dust dominated regions to be locally spatially homogeneous
and isotropic.  These symmetries prevent the sudden formation of
singularities at different points in space, as every point is taken to
be identical to every other point by fiat.  The geometry of these
regions is therefore given by the FLRW line-element (\ref{FLRWds}), where
$a(t)$ is the scale factor
that obeys the Friedmann equation
\be
\frac{\dot{a}^2}{a^2} = \frac{8 \pi G}{3} \rho - \frac{k}{a^2},
\ee
where $\rho \propto a^{-3}$ is the energy density of the dust.  In the
vacuum regions we will take the geometry to exhibit translational
symmetry along the line of sight, but will not restrict the geometry to
be FLRW, as it is known that no solutions that satisfy the junction
conditions will exist in this case.  The geometry of this region is
then given by
\be
\label{kasner}
ds^2 =-d\hat{t}^2 + b_1^2(\hat{t})d\hat{X}^2  +b_2^2(\hat{t})
\left(d\hat{Y}^2+d\hat{Z}^2 \right),
\ee
where $b_1(\hat{t})$ and $b_2(\hat{t})$ are the scale factors in the
directions tangent and normal to the line of sight, respectively.

It can be shown that the junction conditions between these two regions
are satisfied if we identify our hatted and un-hatted coordinates at
the boundary, and if $k=0$, $b_1=a^{-1/2}$ and $b_2=a$ \cite{Lake, Dyer}.
The dust dominated regions are then spatially flat FLRW, the vacuum
regions are Kasner, and the entire geometry is an exact solution of
Einstein's equations \cite{Landry}.  These solutions are, in fact, a
special case of the general dust solution admitting a three
dimensional group of space-like Killing vectors on two dimensional
planar subspaces \cite{Stephani, DiDio}, but are chosen such that we can have
collapsing regions that do not exhibit the shell crossing
singularities that tend to rapidly form in the general case.
We can arrange for either the dust regions to
be collapsing and the vacuum regions to be expanding along the
line of sight, or the dust regions to be expanding and the vacuum
regions to be collapsing along the line of sight.  Here we will
concentrate on the former, which appears to us to be more in-keeping
with the usual picture of what is expected to happen in the late
Universe (voids expanding, and dense regions collapsing).

In order to calculate observational quantities within this solution,
along our chosen line of sight, we will need to specify a set of
observers.  In the dust dominated regions these can be conveniently
chosen to be comoving with the fluid, such that they follow a set of
geodesic curves with tangent vector $u^{a}=(1,0,0,0)$.  In the vacuum
regions we will also take our observers to follow curves with tangent
vector $u^{a}=(1,0,0,0)$, in the coordinates used in
Eq. (\ref{kasner}).  One should note that although there is no fluid
in this case, so that this choice is not unique, it is a choice that
picks out a set of curves that are parallel to the world-lines of
observers who stay at the boundary between
regions.  They are also geodesic.  These choices therefore correspond
to a congruence of complete geodesic curves that fill the entire spacetime.

Let us again refer to the dust dominated regions as Region I and
the vacuum regions as Region II.  Null geodesic curves in these
two regions, in the direction of the inhomogeneity, are given by
\ba
\label{k1}
k_{I}^{a} &=& \frac{c_1}{a^2} \left( a,0,0,-1\right)\\
\label{k2}
k_{II}^{a} &=& c_2 \left( \sqrt{a},0,0,-a \right),
\ea
where $c_1$ and $c_2$ are constants, and where we have again taken the
affine parameter that defines these tangents to decrease into the
past.  The energy of a photon following $k^a$, as measured by the
observers following the curves $u^a$, is given by
\ba
\label{E1}
E_I &=& \frac{c_1}{a}\\
\label{E2}
E_{II} &=& c_2 a^{1/2},
\ea
where $E=-u_a k^a$.  We can also see that the Sachs optical equations
in the two regions reduce to
\ba
\label{eqn-kasner-da1}
\frac{d^2 d_{A,I}}{d\lambda^2} &=& - \frac{2 c_1^2}{3 a^5} d_{A,I}\\
\label{eqn-kasner-da2}
\frac{d^2 d_{A,II}}{d\lambda^2} &=& 0,
\ea
where $c_1$ is the constant from Eq. (\ref{k1}), which will be
different within each individual dust dominated region.  
As always,
the redshift is given by taking the ratio of photon energy at the
time of emission and observation, calculated using Eqs. (\ref{E1}) and
(\ref{E2}).

The trajectories of photons can be straightforwardly integrated between
the different regions, using Eqs. (\ref{k1}) and (\ref{k2}), and
by taking the value of $\hat{X}(\hat{t})$ on leaving one region
as its initial value on entering the next.  Likewise, the value of $E$ can
be calculated along the null trajectories by setting its value on
entering one region as being equal to its value on leaving the last.
This gives the value of the constants $c_1$ and $c_2$ in each of the
dust and vacuum regions respectively, and allows Eqs. (\ref{eqn-kasner-da1}) 
and (\ref{eqn-kasner-da2}) to be integrated along the null trajectory. 
Integration of this equation again requires setting $d_A$ and $\theta$
to be equal at the boundaries between regions.
Following the prescription above, the model is uniquely specified once
we specify three pieces of information:  (i) The size of the vacuum
regions, (ii) the size of the dust regions, and (iii) the time until
the dust regions reach the `big crunch'.

Now let us consider the Buchert average
of this geometry.  The usual procedure is to average the
expansion scalar over a space-like hypersurface, and use the averaged
value to calculate observables.  Here we have created a model that is
inhomogeneous in one direction only.  Averaging in all three spatial
directions should {\it not} therefore be expected to reproduce
anything like the observations we can calculate by looking in the
direction of the inhomogeneity.  Instead, we consider that the appropriate 
thing to do (and in analogy to the case  of inhomogeneity in all 3 spatial 
directions), is to average the scale factor in the direction of inhomogeneity 
only.  We are then left with an averaged geometry with line-element
\be
ds^2 =-dt^2 + \left< b\right>^2 dX^2 + a^2(t) \left(dY^2+dZ^2\right),
\ee
where the averaged scale factor, $\left<b\right>$, is given by
\be
\left<b\right> = \frac{\int \sqrt{g_{XX}} dX}{\int dX}.
\ee
We can now calculate observables in this averaged geometry, and compare them to
the observations made along the line of sight in the actual geometry
of the spacetime.

Let us now consider the spatial average of the local volume 
deceleration parameter. Again, we do not want to
average over all spatial directions, as we are only considering
inhomogeneity and observations along one preferred direction.  The
average of $q_\Theta$ along the line of sight is therefore given by
\be
\left< q_\Theta \right> = \frac{\int q_\Theta \sqrt{g_{XX}} dX}{\int
  \sqrt{g_{XX}} dX},
\ee
where $q_\Theta$ is $1/2$ in the collapsing dust regions, and $-4$ along
the $X$-direction in the vacuum regions.  We find a similar expression
for $q_{\mathrm{KS}}$, which takes the same value as
$q_\Theta$ in both the dust and vacuum regions.

\subsection{Lema\^{i}tre-Tolman-Bondi model}\label{sect-model-ltb}

In the previous two sections, we considered models consisting of alternating 
expanding vacuum and collapsing dust regions. We will now consider a model 
with no discontinuities in the density distribution, in the form of the 
spherically-symmetric, dust-only Lema\^{i}tre--Tolman--Bondi (LTB) solutions 
\cite{LTB1, LTB2, LTB3}. 
These have been the focus of much recent interest due to their ability, in the 
guise of `giant void' models, to reproduce the observed supernova Hubble 
diagram without a cosmological constant \cite{LTBsn1, LTBsn2, Bolejko4, LTBsn4, LTBsn5, LTBsn6, LTBsn7}. 
They are also capable of having an accelerating spatial average under certain 
conditions \cite{Bolejko1, Paranjape, RasanenLTB, Sussman}.

The LTB metric is given by
\begin{equation}
\label{LTBle}
ds^2 = dt^2 - {\frac{a^2_2(t, r)}{(1 - k(r)r^2)}} dr^2 - a_1^2(t, r) r^2 d\Omega^2,
\end{equation}
where $a_2 = (a_1 r)^\prime$ is the radial scale factor. The transverse scale 
factor $a_1$ must satisfy the analogue of the Friedmann equation,
\begin{equation}
\left ( \frac{\dot a_1}{a_1} \right )^2 = \frac{8 \pi G}{3} \frac{m(r)}{a_1^3} - \frac{k(r)}{a_1^2} + \frac{\Lambda}{3}.
\label{LTBFriedmann}
\end{equation}
Primes and over-dots denote partial derivatives with respect to $r$ and $t$, 
respectively. The functions $k(r)$ and $m(r)$ are arbitrary functions of the 
radial coordinate, and may be interpreted as the spatial curvature and a mass 
density at a given radius. The sign of $k(r)$ classifies the differential 
equation, and analytic parametric solutions exist for each sign.
Integrating Eq. (\ref{LTBFriedmann}) with respect to time introduces a third 
arbitrary radial function, $t_B(r)$, which describes the local time since the 
big bang singularity along the world-lines of the dust. 

The metric in Eq. (\ref{LTBle}) is invariant under the transformation 
$r \rightarrow f(r)$, which can therefore be used to set one of the arbitrary functions to 
a simple form, without losing any generality. The LTB solutions are isotropic 
about $r=0$ only, and in general have different expansion 
rates in the radial and transverse directions ($H_1 = \dot{a}_1/a_1$ 
and $H_2 = \dot{a}_2/a_2$, respectively). The density, expansion, and 
shear scalars for this spacetime are
\ba
\rho &=& \frac{(m r^3)^\prime}{3 a_2 a_1^2 r^2} \\
\Theta &=& 2H_1 + H_2 \\
\sigma^2 &=& \frac{1}{3}(H_1 - H_2)^2.
\ea
Observers away from the centre of symmetry have anisotropic distance-redshift 
relations, in general. Rather than solving the Sachs equations in
full for every direction on the sky, we will operate within the 
{\it dipole approximation}\footnote{See the end of
  Section \ref{sect-results-ltb}, and reference \cite{Alnes}, for a
  discussion of the validity of this approximation.}, which assumes
that the dipole term dominates the 
anisotropy of the distance-redshift relation \cite{GBH, Bull, Zibin, Alnes}. 
The dipole is aligned with the radial direction due to the symmetry of the 
model, and so the monopole can be estimated by taking the mean of the angular 
diameter distance in the radial directions facing into and out from the centre 
of symmetry,
\be \label{eqn-da-dipole}
d_A(z)|_{\ell=0} \approx \frac{1}{2}[d_A(+\hat{r}, z) + d_A(-\hat{r}, z)].
\ee
We expect this to be a reasonable approximation for the models considered here. 
The observational deceleration parameter, $q_\mathrm{obs}$, is defined using 
the monopole of $d_A$ only. 

For light propagation purely in the radial direction, 
the tangent vector to the null geodesics is
\be
k^a =\left( (1 + z) , \pm \frac{\sqrt{1 - kr^2}}{a_2} (1+z) ,0 ,0
\right),
\ee
and the redshift is given by integrating
\be
\frac{dz}{d\lambda} = -H_2 (1+z)^2.
\ee
The angular diameter distance in either direction can be found using the 
Sachs equation (\ref{eqn-sachs-theta2}) in the radial direction.

In general, spatial averages in an LTB spacetime will be both position- and 
domain-dependent. We consider spatial averaging only for spherical domains 
centred at $r=0$, on hypersurfaces of constant $t$. We define the
effective scale factor to be
$a_\mathcal{D} = (V_\mathcal{D}/V_{\mathcal{D},0})^\frac{1}{3}$, where
\be
V_\mathcal{D} = 4 \pi \int_0^{r_\mathcal{D}} \frac{a_2 a_1^2 r^2}{\sqrt{1 - k(r) r^2}} dr
\ee
and $r_\mathcal{D}$ is the radius of the spherical domain. The average of a 
scalar quantity is then given by
\be
\langle S \rangle = \frac{4 \pi}{V_\mathcal{D}} \int_0^{r_\mathcal{D}}
S(r, t) \frac{a_2 a_1^2 r^2}{\sqrt{1 - k(r) r^2}} dr.
\ee
We consider only the class of LTB models with $\Lambda = 0$. This means that 
the spacetime is locally decelerating everywhere, $q_\Theta \ge 0$. The 
Kristian-Sachs measure is also necessarily decelerating 
($q_{\mathrm{KS}} \ge 0$).

Nevertheless, the freedom in the radial profiles makes it possible to 
construct models in which the Hubble diagram exactly matches that of an 
accelerating FLRW model for $z>0$, as seen by an observer at the centre of 
symmetry (although see \cite{Romano2}). As a result, LTB models have been 
studied extensively as a 
conventional relativistic explanation of the apparent acceleration inferred 
from supernova observations which does not require the existence of an exotic 
dark energy component, or modifications to the theory of gravity (e.g. 
\cite{LTBsn1, LTBsn2, CFL, GBH2, Bolejko2, Bolejko3, Biswas, YNS, Bolejko4,
  Nadathur, Bull}). Despite some success, however, they ultimately seem unable 
to account for certain combined sets of cosmological 
observables\footnote{For an alternative perspective, 
see \cite{ClarksonRegis}.} \cite{Bull}.

Acceleration of the Buchert average, $q_\mathcal{D} < 0$, has been demonstrated 
for a number of different LTB models (e.g. \cite{Bolejko1, Chuang, Paranjape, RasanenLTB, Kai}). The cases studied generally take 
spherical averaging domains, centred about the origin. They typically find 
acceleration only for finite ranges of domain size, which do not correspond to 
the homogeneity scale (if one exists). Additionally, models in which 
observers at the centre of symmetry infer acceleration from their Hubble 
diagram seem not to correspond to those with an accelerating Buchert average \cite{Bolejko1}, and vice versa \cite{Romano1}. Since the spacetime is always 
decelerating locally, the existence of acceleration in the Buchert average must 
be caused purely by the backreaction term, Eq. (\ref{eqn-Q}). 

In what follows, we will specialise to LTB models with a simple Gaussian 
spatial curvature profile
\be\label{eqn-ltb-profile}
k(r) = A_k \exp \left ( -\frac{r^2}{w_k^2}\right ),
\ee
with $t_B =$constant, and with a choice of radial coordinate such that 
$m(r)=$constant. When $A_k < 0$, there is a central void region 
(often surrounded by an over-dense shell), and an asymptotic flat FLRW region. 
These models are capable of producing good fits to the existing supernova
data, for an observer at the centre of symmetry.

\begin{figure}[b]
\vspace{-0.5cm}
  \includegraphics[width=9.5cm]{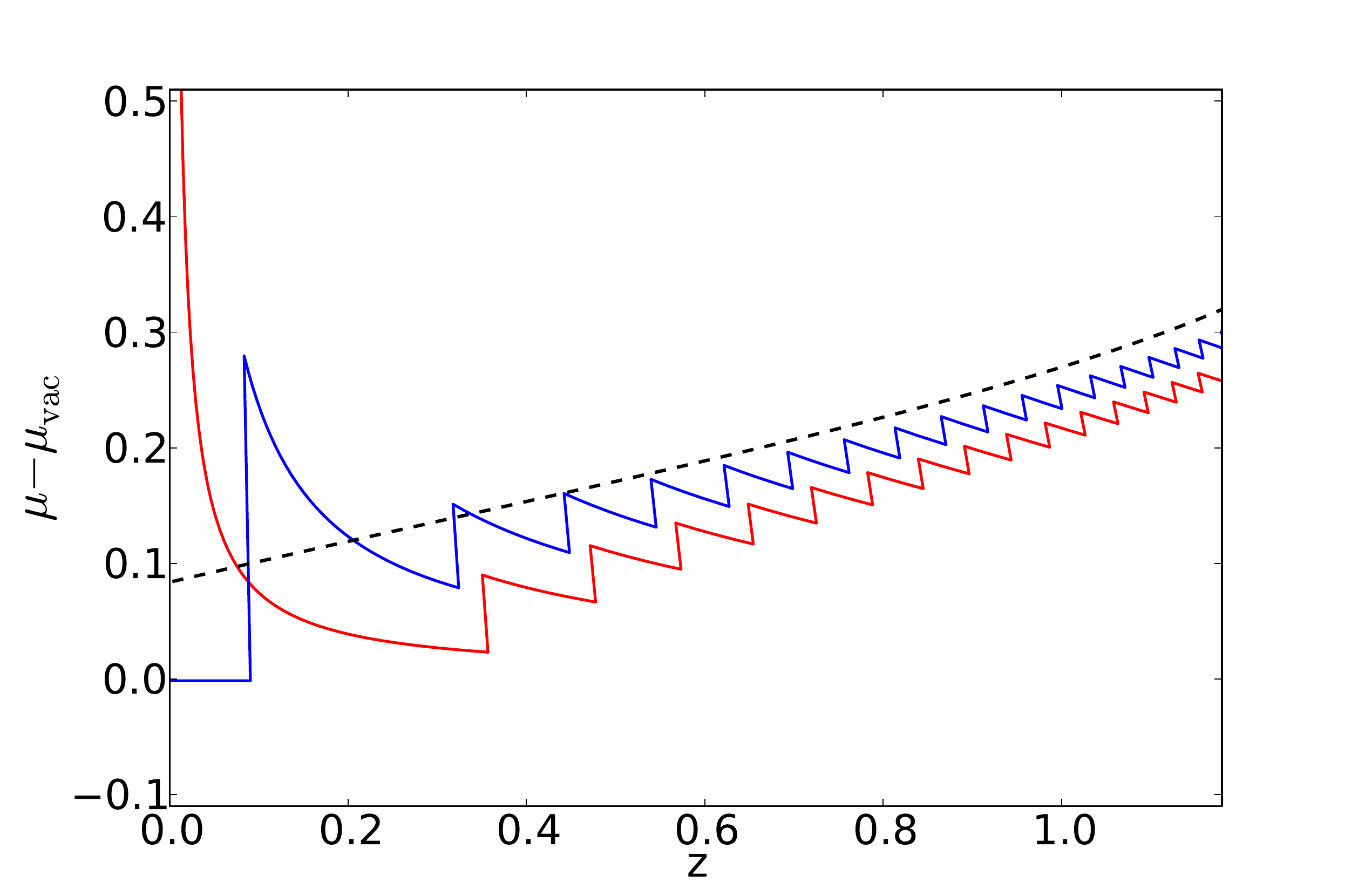}
\vspace{-0.5cm}
 \caption{Magnitude of sources in the spherical collapse model, minus the
magnitude they would have in pure vacuum.  Observations made by a single 
observer, from the centre of the dust and vacuum regions, are displayed as 
solid red (lower at $z=1$) and blue lines, respectively.
The dashed  black line is the same quantity in the Buchert averaged model.}
 \label{fig-sc3-1}
\end{figure}

%----------------------------------------------
\section{Results: Spherical collapse}\label{sect-results-sc}
%----------------------------------------------

\begin{figure*}
\vspace{-1cm}
  \includegraphics[width=19cm]{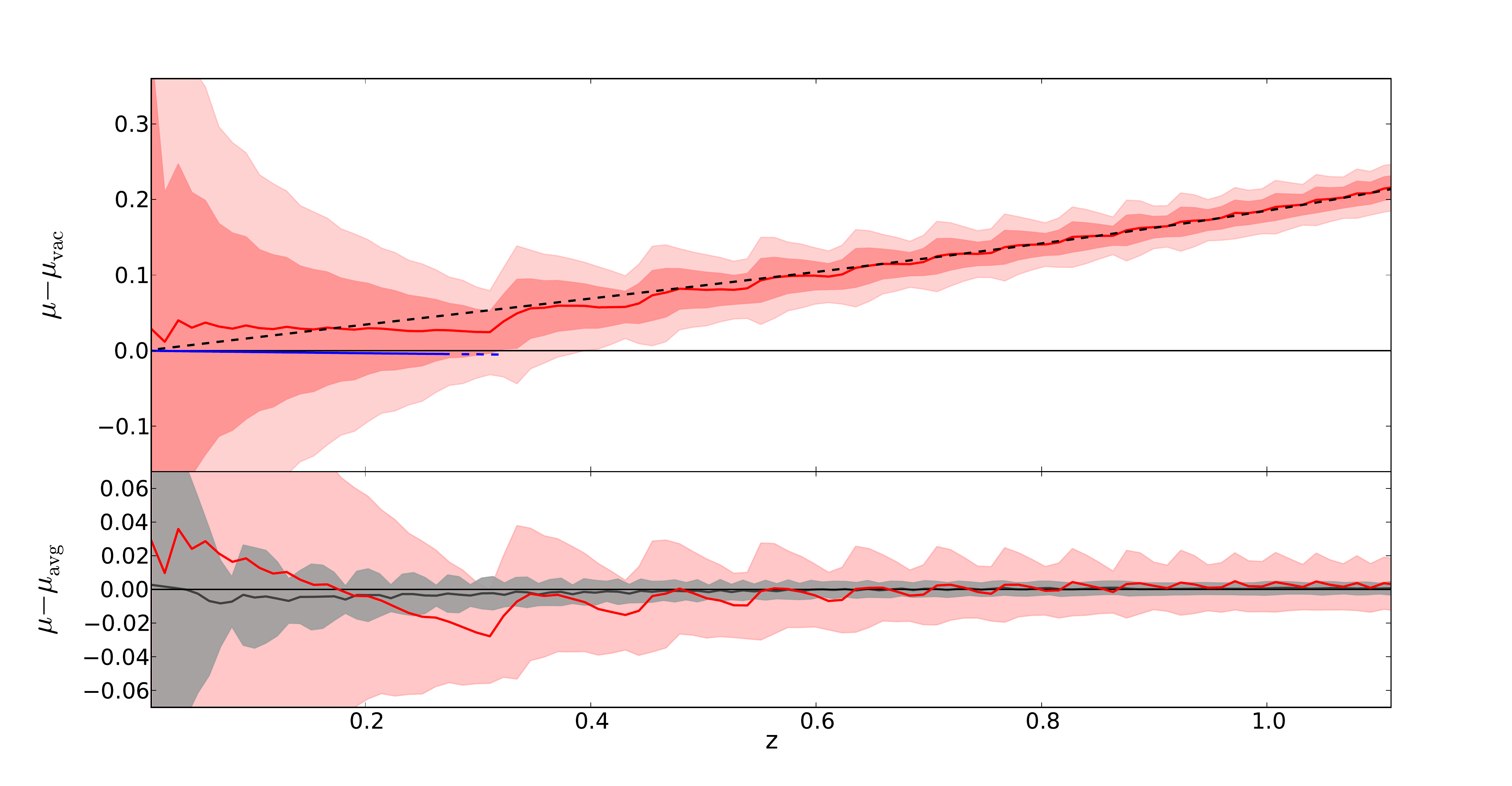}
\vspace{-0.5cm}
 \caption{Upper Panel: 
Magnitude of sources in the spherical collapse model, minus the
magnitude they would have in pure vacuum.  The solid red line is the
mean observed magnitude, 
obtained by spatially averaging distance-redshift relations on a surface of
$t=$constant. The pink bands are $1\sigma$ and $2\sigma$ confidence regions. 
The black dashed line shows the corresponding quantity in the Buchert averaged model. 
The short blue line is the distance modulus for a FLRW model with deceleration parameter 
$q = \langle q_\Theta\rangle$.
Lower Panel: A difference plot of the red line and dashed line, from
the panel above.  The solid black line corresponds 
to a model with regions that are a quarter of the size of those used
in the upper panel.  The pink (wide) and grey (narrow) bands are
$1\sigma$ confidence regions for the two different models.}
 \label{fig-sc3-23}
\end{figure*}

\begin{figure*}
 \centerline{
  \includegraphics[height=10cm]{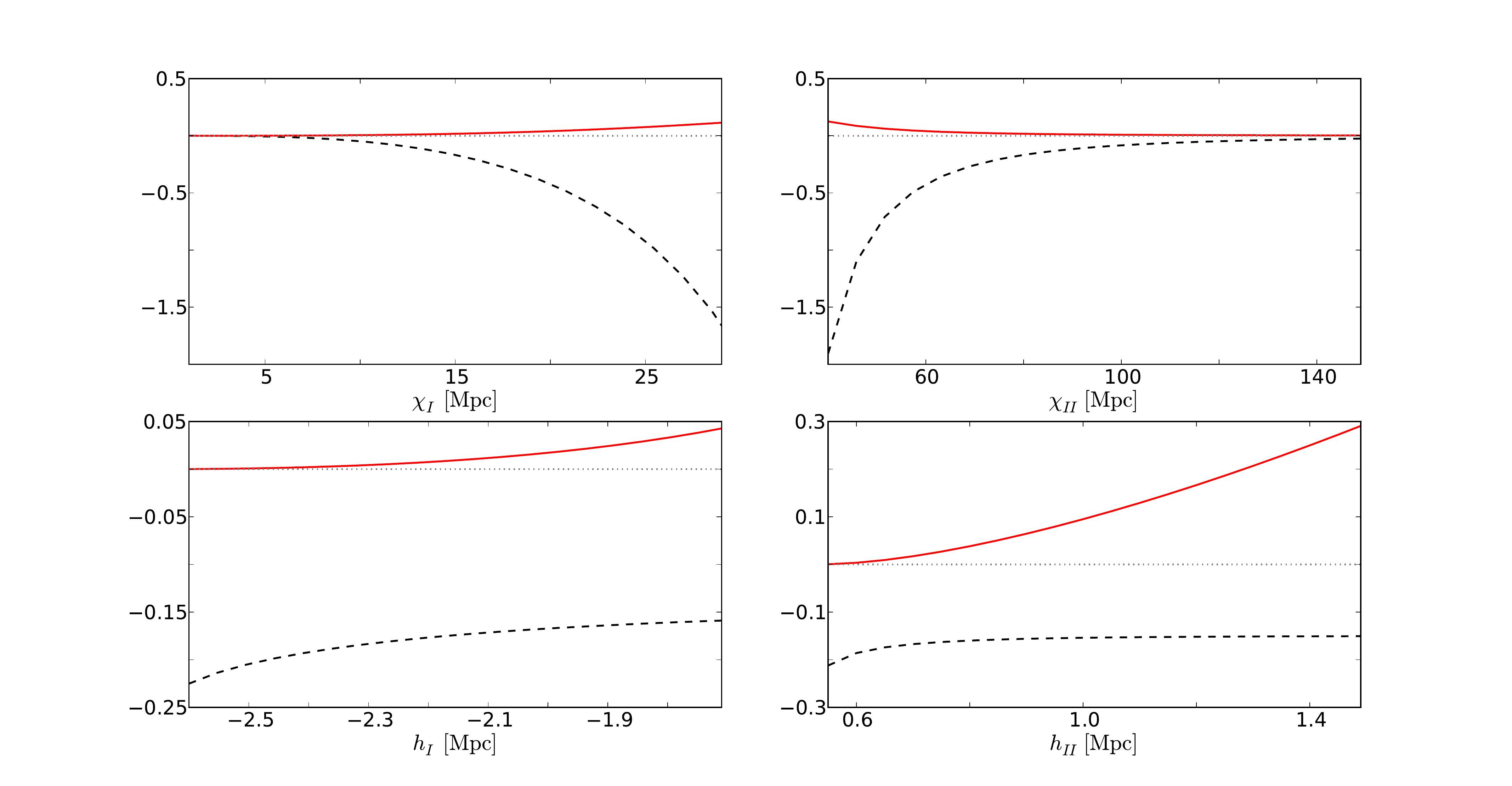} }
 \caption{Dependence of the Buchert average deceleration parameter, $q_\mathcal{D}$ (dashed black 
 line), and spatially-averaged local volume deceleration parameter, $\langle q_\Theta \rangle$ (solid 
 red line), on the parameters of the spherical 
 collapse model. Shown is the dependence on the Hubble rates in the collapsing 
 and vacuum regions, $h_I$ and $h_{II}$, and the comoving region
 sizes, $\chi_I$ and $\chi_{II}$.}
 \label{fig-sc3-4}
\end{figure*}

In this section, we evaluate our 4 measures of acceleration in the 
spherical collapse model (as described in Section \ref{sect-model-sc}). This model 
is known to be able to have an accelerating Buchert average, despite locally decelerating 
everywhere \cite{IshibashiWald}.

In Figure \ref{fig-sc3-1} we plot the magnitude of sources, $\mu$, that an
observer in the spherical collapse model would see, minus the
magnitude a source at the same redshift would have in a pure vacuum
model (with all of the dust regions removed).  The model used in this
figure is one in which the vacuum regions are $80$ Mpc wide at the
present time, and dust regions are $15$ Mpc wide. We refer to these
plots as `the Hubble diagram'.  We consider two different observers in this plot:
one at the centre of a collapsing dust region (red line), and one at the centre of an 
expanding vacuum region (blue line). It can be seen that the type of
region that the observer finds themselves in can have a considerable
impact on the behaviour of the Hubble diagram they construct at low
$z$.  At high $z$, however, the curves evolve almost 
identically (up to a vertical displacement). In both cases, the
blueshifting that occurs in the collapsing region forces the curves 
upwards, due to the magnitude continuing to increase even while the redshift 
decreases.  This leads to an overall positive gradient for the curves
in Figure \ref{fig-sc3-1}, while the gradient at any point on
each individual curve is negative.  This result is significant, as
accelerating FLRW models have positive gradient in these plots, while
decelerating FLRW models have negative gradient.

In Figure \ref{fig-sc3-23} we plot the results of averaging the Hubble
diagrams constructed by observers in both the dust and vacuum
regions of the same model considered in Figure \ref{fig-sc3-1}.  We do
this by constructing individual distance-redshift relations for a
large number of observers (all in the same model), binning these relations in 
redshift,
and then calculating the mean and standard deviation in each bin.
We also calculate the magnitudes that would be found in the
Buchert averaged model, and the spatially averaged deceleration
parameter $q_\Theta = q_{\mathrm{KS}}$.  In the upper panel of Figure
\ref{fig-sc3-23} we plot the magnitude of sources, minus the magnitude
they would have at the same redshift if all the dust regions were
removed (as in Figure \ref{fig-sc3-1}).  In the lower panel we plot
the magnitude of sources minus the magnitude they would have in the
Buchert averaged model.  In this figure we shift the curves so that
they coincide at $z=0$.  This corresponds to a change in the local Hubble rate.

It can be seen from the upper panel of Figure \ref{fig-sc3-23} that the Buchert average (dashed black 
line) closely traces the mean observed magnitude (solid red line). 
The $1$ and $2\sigma$ confidence regions appear to oscillate, because of the jaggedness of the individual 
distance-redshift curves that were averaged over (see Figure 
\ref{fig-sc3-1}).  Nevertheless, the curve for the Buchert average stays within the $1\sigma$ 
confidence region for the entire redshift range considered.
Part of the reason for this is that the 
black line has been shifted vertically to match the red line at high $z$,  
in order to aid comparison. Without this shift, the two curves 
have the same zero-point, but fluctuations in the mean observational curve at 
low $z$ cause an offset to build up at higher $z$. The zero-points are the 
same because they are governed by the same spatial average of the local Hubble rate 
$\langle H_0 \rangle$ (for the observational curve) and the Buchert average 
Hubble rate $H_\mathcal{D}$ (for the Buchert average curve).  These are equal at 
$z=0$.

It can be seen from the lower panel in Figure \ref{fig-sc3-23} that making the comoving size of the 
regions smaller, so that the initial region is less dominant, significantly 
reduces the fluctuations of the mean observational curve (in both cases, these
fluctuations decrease as redshift increases). This in turn reduces 
the offset that develops between the observational curve and the Buchert 
average curve.  

The short blue line in the upper panel of Figure \ref{fig-sc3-23} is the `effective'
curve obtained by setting $q_0 = \langle q_\Theta \rangle$ in the 
FLRW series expansion for $d_A(z)$. Again, $q_\Theta = q_{\mathrm{KS}}$ 
in the spherical collapse model, so this line is also the effective curve for 
the Kristian-Sachs deceleration parameter. It can be seen to bear little resemblance to the 
observational and Buchert average curves.  

The values for the deceleration parameters measured in this model are
\begin{eqnarray}
\langle q_\Theta \rangle &\simeq& 0.017\\
q_\mathcal{D} &\simeq& -0.167,
\end{eqnarray}
with $\langle q_{\mathrm{KS}} \rangle =\langle q_\Theta \rangle$, and
$\langle q_{\rm obs} \rangle \simeq q_\mathcal{D}$.
By considering other model parameters, we have confirmed that
the curves due to averaging observations, and the Buchert average distance curve, 
always seem to show similar functional behaviour (and thus have similar deceleration 
parameters).  Neither of these quantities are ever close to $\langle
q_{\mathrm{KS}} \rangle$ or $\langle q_\Theta \rangle$, unless the
dust regions are made to expand, or are made to be very small.

We find that the existence of acceleration in both the Buchert average
and the observational 
distance-redshift relations is a generic feature of spherical collapse 
models consisting of expanding and collapsing regions. Figure \ref{fig-sc3-4} 
shows the results of varying the parameters of the FLRW regions that constitute 
the model. The base model has 
the parameters $h_I = -2.0$ and $\Omega_{m,I} = 1.8$ in the dust
regions, and $h_{II} = 0.7$ and $\Omega_{m,II} = 0$ in the vacuum
regions.  The comoving sizes of these regions are taken to be $\chi_I = 15$ 
Mpc and $\chi_{II} = 80$ Mpc, respectively. The collapsing 
region is chosen to be 52\% of its maximum age. 
The plots in Figure \ref{fig-sc3-4} are for models with these
parameter values, unless they are the parameter being varied. The 
figure shows a strong dependence of the Buchert average acceleration on region 
size -- models with collapsing regions that are relatively larger have greater 
acceleration, as expected. Increasing the Hubble rate in the vacuum regions reduces the 
current age of the model, and correspondingly increases $\langle q_\Theta \rangle$.
This is due to this quantity being evaluated at an earlier time, when the collapsing regions 
are more dominant. Figure \ref{fig-sc3-4} shows that the deceleration
parameter of the Buchert averaged model is 
negative as long as the collapsing region takes up a non-negligible fraction 
of the comoving volume. If the vacuum region dominates, both the local volume 
and Buchert average deceleration parameters tend to zero. Since the 
observational deceleration parameter is well-approximated by the Buchert average 
measure, it seems that all that is required for observers to see an apparent 
acceleration is the existence of a non-negligible fraction of collapsing regions.

%----------------------------------------------
\section{Results: Kasner-EdS}\label{sect-results-kasner}
%----------------------------------------------

\begin{figure}[b]
 \centerline{
  \includegraphics[width=9.5cm]{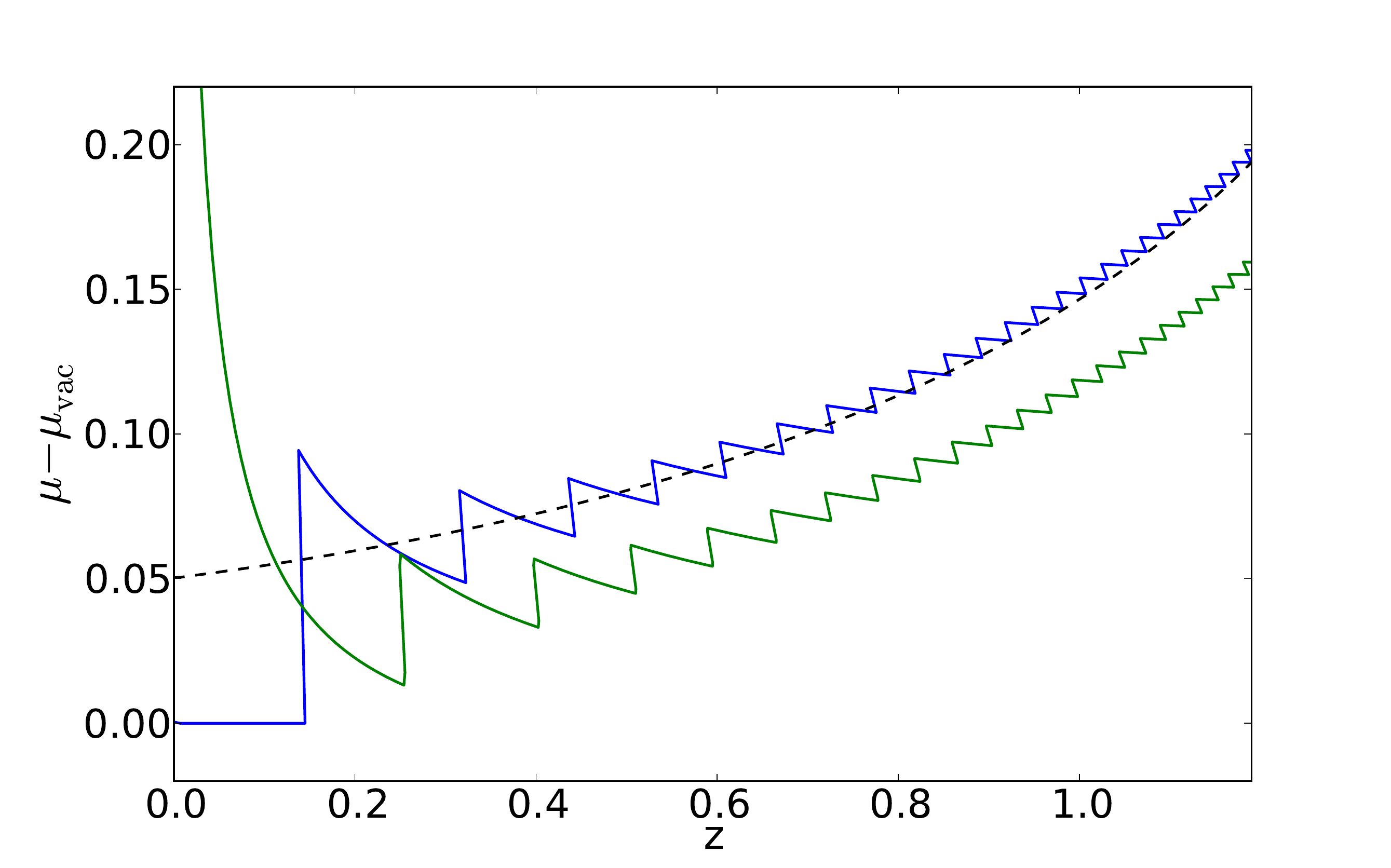} }
 \caption{The same quantities that were plotted in Figure \ref{fig-sc3-1}, but
  using the Kasner-EdS model.}
 \label{fig-kasner-1}
\end{figure}

\begin{figure*}
\vspace{-1cm}
\includegraphics[width=19cm]{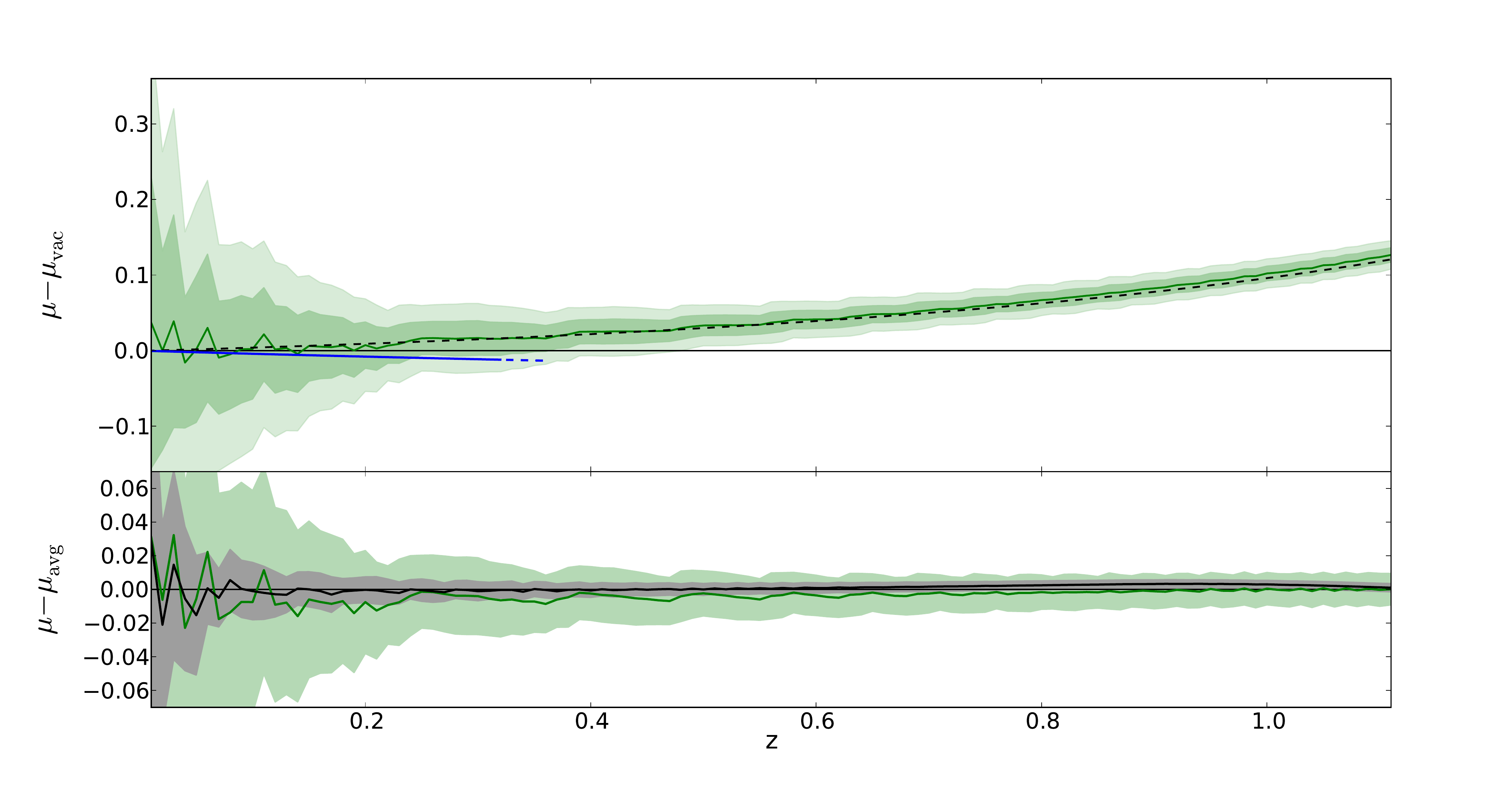}
\vspace{-0.5cm}
\caption{The same quantities that were plotted in Figure \ref{fig-sc3-23}, but
  using the Kasner-EdS model.}
    \label{fig-kasner-23}
\end{figure*}

We now repeat the analysis of the previous section for the Kasner-EdS
model. This is included in order to allay concerns that the coherence in the 
accelerations of the Buchert average and observational measures seen in the 
spherical collapse model were caused simply by inadequacy of the model.

In Figure \ref{fig-kasner-1} we plot the magnitude of sources, $\mu$, that an
observer in the Kasner-EdS model would see, minus the
magnitude a source at the same redshift would have in a pure vacuum
model (with all of the dust regions removed).  The model used in this
figure again has dust regions that are $15$ Mpc wide, but the vacuum
regions are now $\sim 1500$ Mpc across.  This large number is required
in order to get cosmologically interesting redshifts, because of the rapid
acceleration in the vacuum regions along the preferred line of
sight\footnote{It  should be reiterated here that we are not proposing
  this model as a realistic 
representation of the actual Universe, but rather to provide a simple exact
solution of Einstein's equations that allows us to study relativistic
behaviour that may occur in more realistic solutions. In this sense,
it should be considered as proof-of-concept only.}.
We also choose the model such that the dust regions collapse to a
singularity in $\sim 5$ billion years.
We consider two different observers in Figure \ref{fig-kasner-1}:
one at the centre of a collapsing dust region (green line), and one at the centre of an 
expanding vacuum region (blue line). Again, it is apparent that the type of
region that the observer finds themselves in can have a considerable
impact on the behaviour of the Hubble diagram they construct at low
$z$, while at high $z$ the curves evolve almost 
identically (up to a vertical displacement). Blueshifting again
occurs in the collapsing regions, forcing the curves upward.

In Figure \ref{fig-kasner-23} we plot the results of averaging the 
distance-redshift relations of observers in both the dust and vacuum
regions of the same model considered in Figure \ref{fig-kasner-1}, as
was done in Figure \ref{fig-sc3-23} for the spherical collapse model.
Magnitudes in the Buchert averaged model, and the spatially averaged deceleration
parameter $q_\Theta = q_{\mathrm{KS}}$, are also calculated.  With this
model, however, no vertical shift of the Buchert averaged model curve is required in
order for it to agree well with the average of the observed
magnitudes. The Buchert average closely traces the mean observed
magnitudes. We do, however, perform a vertical shift on all of the
curves in Figure \ref{fig-kasner-23} simultaneously, so that they approach the 
origin at $z=0$. Again, reducing the comoving size of the 
regions, so that the initial region is less dominant, significantly 
reduces the fluctuations of the mean observational curve.

The short blue line in the upper panel of Figure \ref{fig-kasner-23} is
again the `effective' distance 
modulus curve obtained by setting $q_0 = \langle q_\Theta \rangle$ in the 
FLRW series expansion for $d_A(z)$. We also again have $q_\Theta = q_{\mathrm{KS}}$.
The values for the deceleration parameters measured in this model are
\begin{eqnarray}
\langle q_\Theta \rangle &\simeq& -3.96\\
q_\mathcal{D} &\simeq& -7.18,
\end{eqnarray}
with $\langle q_{\mathrm{KS}} \rangle =\langle q_\Theta \rangle$, and
$\langle q_{\rm obs} \rangle \simeq q_\mathcal{D}$.
This value of $\langle q_\Theta \rangle$ corresponds to rapid
acceleration, but not as rapid as simply looking through pure vacuum
regions in the direction of inhomogeneity, where $q_0 = -4$.  The
value of $q_\mathcal{D}$, on the other hand, corresponds to
considerably more acceleration than simply looking through the vacuum
regions alone.  As in the spherical collapse models, therefore, the
presence of the collapsing dust regions causes a dramatic increase in
both the acceleration of the Buchert averaged model, and the average
of the observational acceleration.

By considering models with other parameter values, we again confirm that
the curves due to averaging observations, and the Buchert average distance curve, 
always seem to show similar functional behaviour (and thus have similar deceleration 
parameters).  Neither of these quantities are ever close to $\langle
q_{\mathrm{KS}} \rangle$ or $\langle q_\Theta \rangle$, unless the
dust regions are removed, or become vanishingly small.

%----------------------------------------------
\section{Results: LTB}\label{sect-results-ltb}
%----------------------------------------------

\begin{figure*}
\includegraphics[width=17cm]{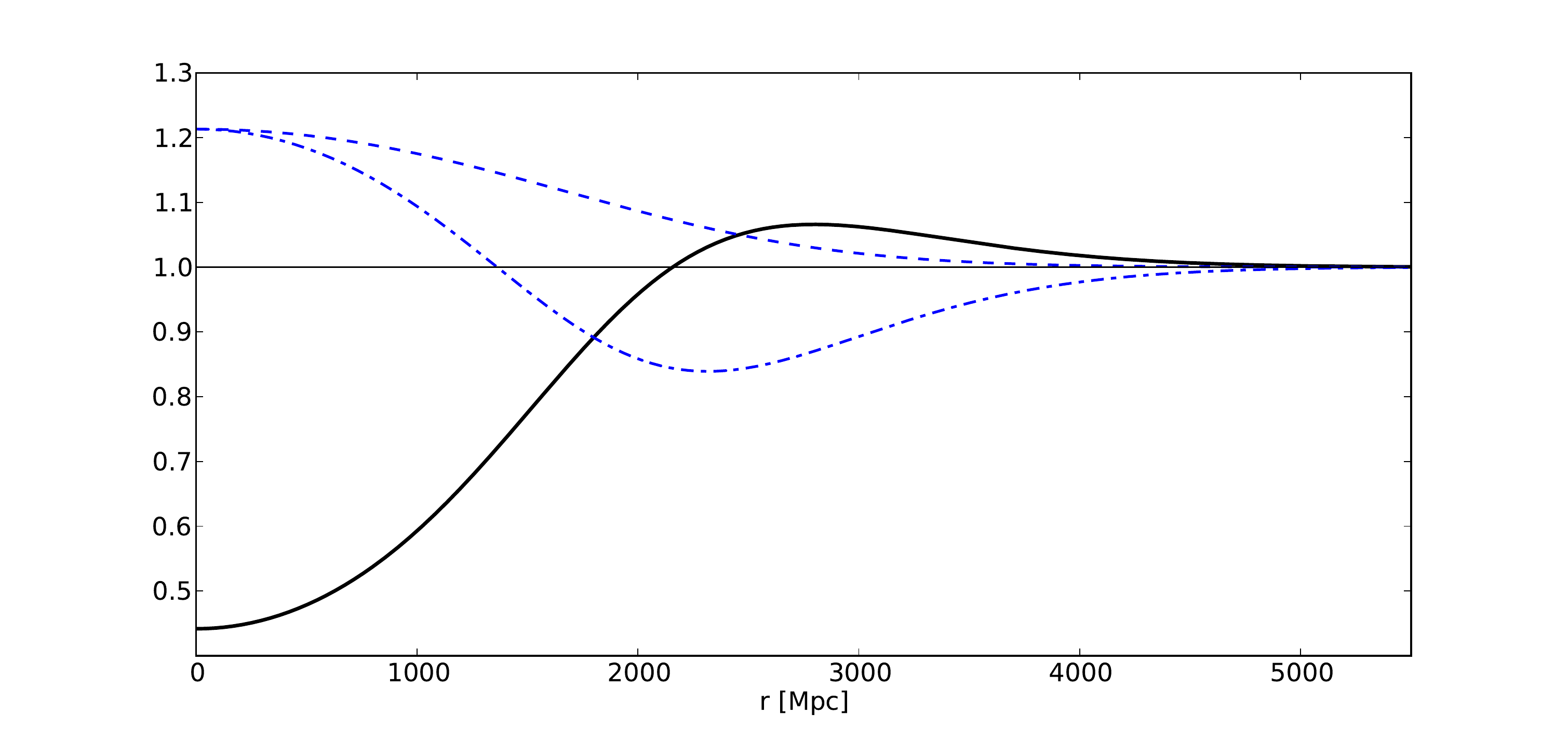}
\caption{Density and expansion rates as a function of radius in an 
 LTB void model. The model has spatial curvature of the form given in 
 Eq. (\ref{eqn-ltb-profile}), with $A_k=-3.82\times 10^{-8} \mathrm{Mpc}^{-2}$ 
 and $w_k = 1800 ~\mathrm{Mpc}$. Shown are the density, $\rho$ (solid black), 
 transverse Hubble rate, $H_1$ (blue dashed), and radial Hubble rate, $H_2$ 
 (blue dash-dot). All of the curves are normalised to their values in the 
 asymptotically-homogeneous region 
 ($\rho = 9.2\times 10^{10} \mathrm{M}_\odot \mathrm{Mpc}^{-3}$, 
 $H_1 = H_2 = 57.7$
 kms$^{-1}$Mpc$^{-1}$).}
\label{fig-ltb-1}
\end{figure*}

\begin{figure*}
\includegraphics[width=17cm]{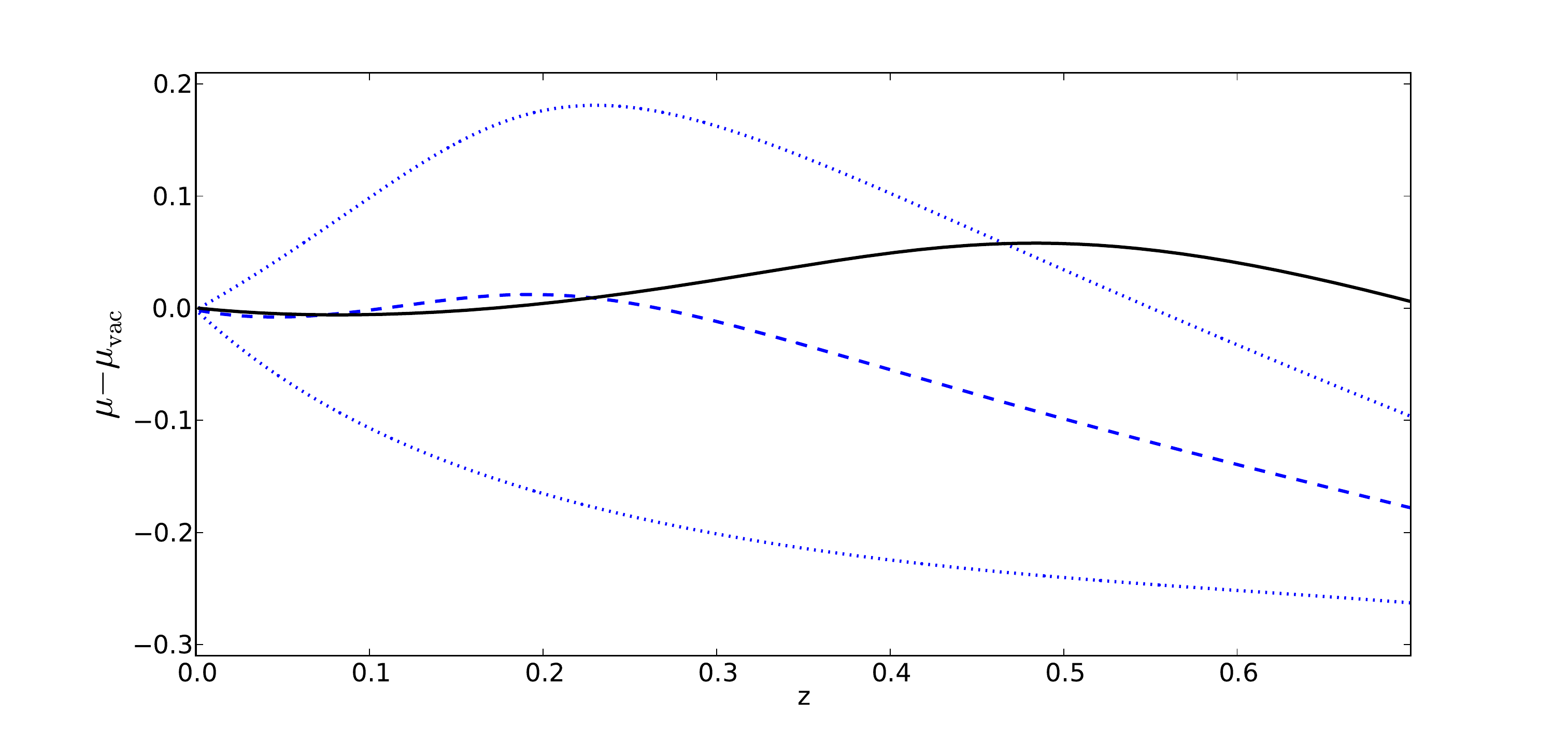}
\caption{Distance modulus curves for an off-centre observer in the same 
  model, at  $r = 915 ~\mathrm{Mpc}$. The dotted blue lines show the distance modulus as a function of 
 redshift for radial lines of sight looking out of and into the void (upper and 
 lower curves respectively). The dashed blue line is the distance modulus for the 
 monopole of the distance-redshift relation, calculated using the dipole 
 approximation. The solid black line is the relation for an observer at the 
 centre of symmetry.}
\label{fig-ltb-2}
\end{figure*}

\begin{figure*}
 \centerline{
  \includegraphics[width=17cm]{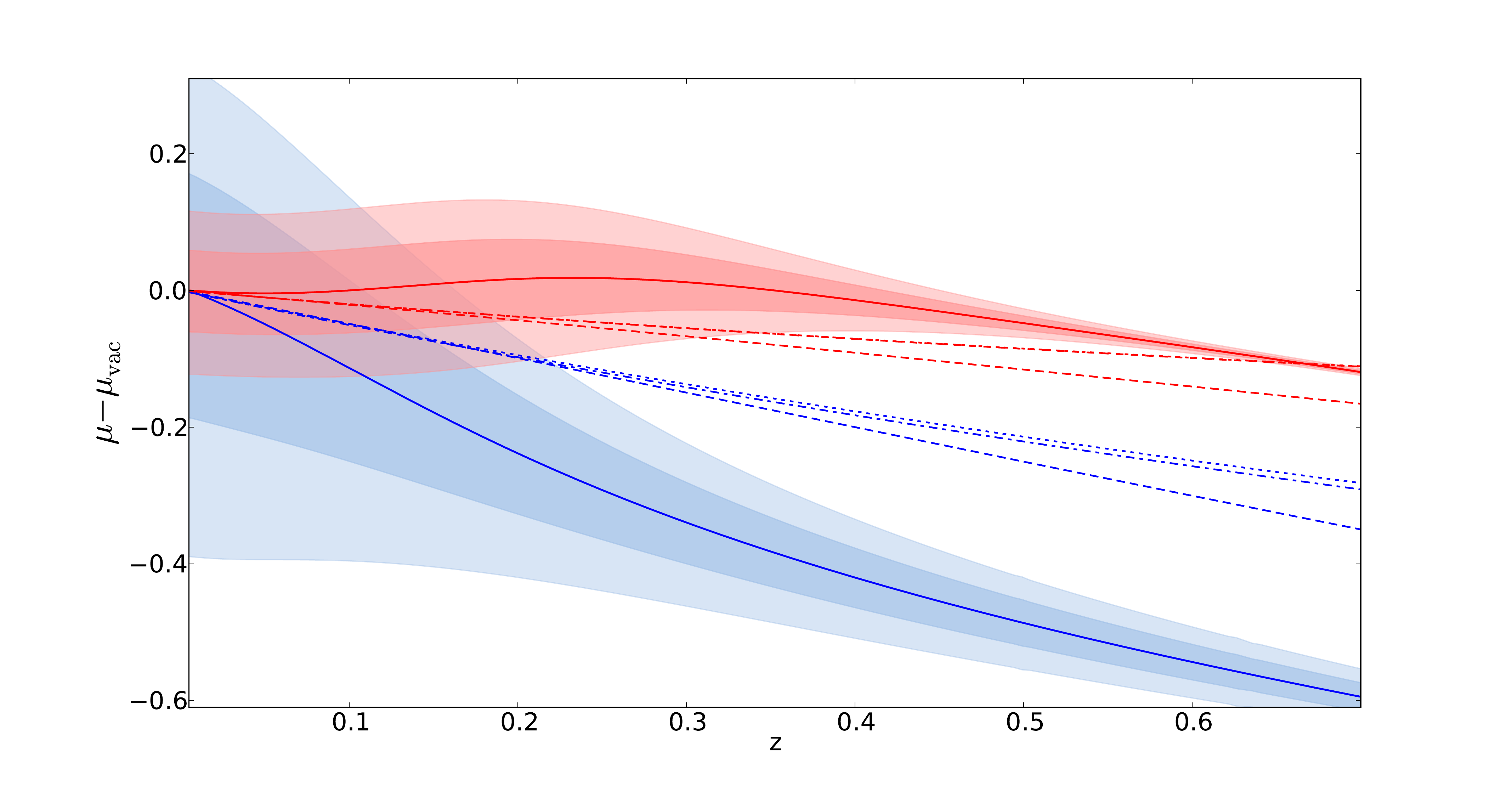} }
 \caption{Distance moduli in the LTB model. The red curves (uppermost) are for 
 averaging domains of size $r_\mathcal{D}=1000$ Mpc, and the blue curves for 
 $r_\mathcal{D}=3000$ Mpc. The solid lines are the mean distance moduli found 
 by spatially averaging the monopole of the observed distance-redshift 
 relation, with 
 accompanying $1\sigma$ and $2\sigma$ confidence bands. The dashed lines are 
 the distance moduli for the distance-redshift relation in the Buchert averaged 
 spacetime. The dotted and dot-dashed lines are the distance moduli for a 
 series expansion of the FLRW $d_A(z)$ relation, with deceleration parameters 
 $q = \langle q_\Theta\rangle$ and $q = \langle q_{\mathrm{KS}}\rangle$, 
 respectively. The curves have all had their zero-points shifted to match 
 at $z=0$.}
 \label{fig-ltb-3}
\end{figure*}

\begin{figure*}
 \centerline{
  \includegraphics[width=17cm]{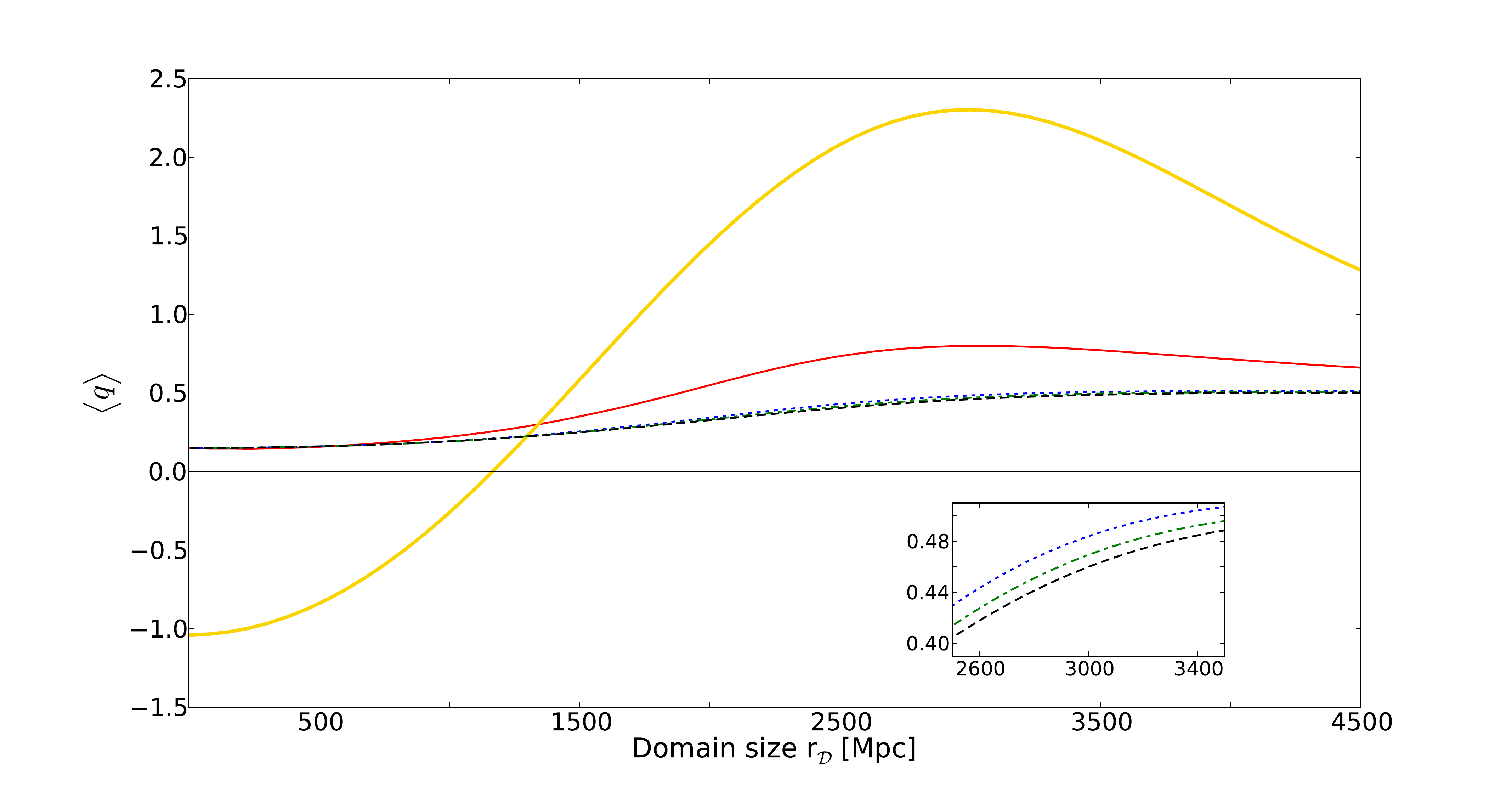} }
 \caption{Spatial averages of various deceleration parameters as a function of 
 averaging domain radius, $r_\mathcal{D}$. The local volume (dash-dotted green 
 line), Kristian-Sachs (dotted blue), and Buchert average (dashed black) 
 deceleration parameters have rather similar values. A close-up of these curves 
 is shown in the inset so that they can be distinguished. The thick, solid 
 yellow line is the deceleration parameter found by fitting an FLRW model to 
 the distance-redshift relation for a wide range of redshifts, $z\le 1$. The 
 narrower solid red line shows the same measure, but for a fit to only 
 $z\le 0.1$.}
 \label{fig-ltb-4}
\end{figure*}

The final model that we consider is an LTB spacetime, with parameters chosen to 
produce a good fit to the supernova data for an observer at $r=0$.
These are displayed in Figure \ref{fig-ltb-1}. 
Unlike the other models we have considered, the LTB model has smooth
density and Hubble rate profiles on spatial 
slices. There are, however, no collapsing regions at
$t=t_0$.

Figure \ref{fig-ltb-3} displays the magnitudes of sources in the LTB model for our various 
measures of acceleration. Since the chosen model has no homogeneity scale, 
we plot the distance modulus curves for two different averaging domain radii: 
$r_\mathcal{D} = 1000$ Mpc (well inside the void), and $r_\mathcal{D} = 3000$ 
Mpc (near the void boundary). The results in the two cases differ significantly, 
with the smaller averaging domain displaying acceleration for the mean 
observational curve (solid red line), while its counterpart for the larger 
domain (solid blue line) shows a strong deceleration. Note that the 
observational curves are for the {\it monopoles} of the distance-redshift 
relation, obtained using the dipole approximation described in Section 
\ref{sect-model-ltb}.
The curves for the Buchert average (dashed lines) do not match the mean 
observational curves (solid lines) in any discernible way. In particular, both 
of the curves for the Buchert average are decelerating, while the averaged 
observed relation for the smaller domain size is accelerating.

The dotted and dash-dotted lines are `effective' distance moduli for the local 
volume and Kristian-Sachs deceleration parameters. These correspond to series 
expansions of the FLRW $d_A(z)$ relation with deceleration parameters 
$q_0 = \langle q_\Theta\rangle$ and $q_0 = \langle q_{\mathrm{KS}}\rangle$ 
respectively. Neither follows the Buchert average, nor the mean observational curves. 
They are closely related to one another though, with only a slight 
discrepancy apparent at higher redshift for the larger domain size. This is 
because the two measures only differ by a term proportional to the shear, 
which is relatively small in this case.

The upper set of observational curves in Figure \ref{fig-ltb-3} show that a 
substantial fraction of observers within a radius of $r \le 1000$ Mpc will 
infer acceleration from the monopole of their Hubble diagram, despite all of 
the other deceleration parameters indicating deceleration. This is caused by 
spatial variations in expansion rate along their lines of sight, but does not 
seem to be linked with spatial average properties of the spacetime, as it was 
in the spherical collapse and Kasner-EdS models.
Fig. \ref{fig-ltb-2} suggests that many of these observers will see a large 
dipole in the distance-redshift relation over their sky. Presumably, most of 
the observers would conclude that they lived in a Universe that is 
inhomogeneous on large scales, and would therefore not attempt to fit an FLRW model to 
their observations at all. In that case, the question of whether the monopole of 
the observational relation is accelerating or not becomes less of an issue. 
Attempts to summarise the (observational) acceleration of the model with one 
number, the monopole $q_{\mathrm{obs}}$, would fail. Attempting 
to construct a homogeneous Buchert average model would probably not be seen as a sensible 
procedure either\footnote{Instead, the Buchert averaging procedure could be 
used to define a `smoothed-out' model that is inhomogeneous. This would result 
in position-dependent average quantities, and would likely be
sensitive to both the shape 
and size of the averaging domain.}.

The Buchert average curve in Figure \ref{fig-ltb-3} is much closer to the 
effective local volume acceleration and Kristian-Sachs curves than to the 
observational curve, although it still deviates from both of them 
substantially. There are a number of reasons why this is the case. The 
first is that the model is not statistically homogeneous, and so the Buchert 
average does not represent the `typical' conditions that a light ray would 
experience traveling through the spacetime\footnote{In fact, it
  cannot, as the null geodesics involved in calculating this quantity
  go outside of the averaging domain.}. Secondly, the backreaction scalar 
$\mathcal{Q}_\mathcal{D}$, of Eq. (\ref{eqn-Q}), is small in this model, because 
both the shear and the variance of the expansion rate are small. As 
such, the Buchert average deceleration parameter $q_\mathcal{D}$, defined in 
Eq. (\ref{eqn-qd}), is 
dominated by the same terms that appear in the definitions of the 
Kristian-Sachs and local volume measures, most notably the density. Our 
different measures of acceleration do not {\it exactly} reduce to one another in this case, but they 
are rather close (as can be seen in Figure \ref{fig-ltb-4}, inset). 
We also see that in the absence of a sensible way of 
defining a representative smooth model (due to the lack of a homogeneity 
scale), the mean behaviour of light rays should not be expected to
correspond to the Buchert averaged model.

In Figure \ref{fig-ltb-4}, the various (spatially-averaged)
deceleration parameters are plotted as a function of averaging domain radius, 
$r_\mathcal{D}$. 
The Buchert average, local volume, and Kristian-Sachs deceleration parameters 
track one another rather closely over the whole range of domain sizes, for the 
reasons discussed above. The observational curves (thick and thin solid lines) 
have a rather different behaviour. Both are obtained by calculating $q_0$ in an 
FLRW model that has been fit to the monopole of the distance-redshift relation 
as a function of observer position\footnote{The best-fit FLRW model 
 may not always be a particularly good fit -- the monopole of the distance-redshift 
 relation in LTB models can take on much more complicated functional forms 
 than are allowed in FLRW.}. The resulting position-dependent value of 
$q_0(r)$ is then spatially averaged. The thick yellow curve corresponds to 
fitting FLRW models to the distance-redshift monopole out to $z=1$, and the 
thin red one to $z=0.1$. 
The two are very different.  In particular, the fit out to higher 
redshifts shows acceleration for small averaging domain radii. Figure 
\ref{fig-ltb-2} shows why this is the case: At low redshift, the observational 
distance modulus curve is decelerating (which is also suggested by the fact that 
$q_{KS} > 0$ everywhere), but appears to accelerate at higher redshifts for 
observers inside the void. By fitting FLRW models to the larger redshift range, 
more of the apparent acceleration is captured. This behaviour is wholly due to
quantities being integrated along past null directions, and is not
caused by the local curvature of spacetime at any one point.

The results we have presented for the observational deceleration are
subject to the validity of the
dipole approximation, that was discussed in Section \ref{sect-model-ltb}. 
This approximation will tend to overestimate the monopole of the angular 
diameter distance at a given redshift, causing a decrease in $q_\mathrm{obs}$ 
relative to its exact value. 
As can be seen from Figure \ref{fig-ltb-4}, this only serves to worsen the 
discrepancy between $\langle q_\mathrm{obs} \rangle$ and the other 
deceleration parameters at high redshift. At low redshift the
approximation is most accurate \cite{GBH}, and so the correction
required in this regime should be expected to be small. As such, we 
consider our results to be robust to the use of this approximation.

%----------------------------------------------
\section{Discussion}\label{sect-discussion}
%----------------------------------------------

In this paper, we have studied different concepts of what it means for
a spacetime to display `accelerating expansion'.  The measures of
acceleration that are associated with these concepts all reduce to the same quantity
in a perfectly homogeneous and isotropic FLRW universe, but in an inhomogeneous
universe we have shown that they can be very different indeed.
This occurs even to the extent that some can indicate deceleration, while
others indicate that exactly the same spacetime is accelerating.  In
universes that are statistically homogeneous on large scales, we find, 
that in order to estimate the acceleration inferred by making
observations over large distances (as is the most usual way to infer
acceleration in cosmology), one is best off using a model constructed
from non-local averages of geometric quantities,
as occurs in Buchert's formalism, rather than considering the local
expansion rate of space. This is in agreement with an argument put forward by 
R\"as\"anen \cite{RasanenLP1, RasanenLP2}.
The appearance of acceleration in observations made over large scales does not 
necessarily imply or require the expansion of space to be accelerating, nor 
does it require local observables to indicate acceleration.

The models that we used to reach these conclusions are given in
Section \ref{sect-models}, and include both exact and approximate 
relativistic toy models that are known to display some of the types of 
acceleration we have considered. To be specific, we consider: (A) an
approximate `spherical collapse' model (with disjoint collapsing and
expanding FLRW regions); (B) an exact
Kasner-EdS model (with expanding and collapsing regions along the 
line of sight); and (C) an exact LTB model (expanding 
everywhere).  The spherical collapse model has the advantage of being
able to model reasonably complicated distributions of matter, while
the Kasner-EdS model allows one to model a universe that is
statistically homogeneous along the line of sight. 

In the spherical collapse and Kasner-EdS models, the reconstructed distance-redshift 
relation, which corresponds most closely to what is actually measured
in observational cosmology, is closely related to the Buchert average, and not the mean 
local properties of the spacetime.  This means that showing that 
local spacetime cannot accelerate without $\Lambda$, or a
quintessence field, is not sufficient to disprove backreaction as a
source for the apparent late-time accelerating expansion of the Universe 
\cite{GreenWald, IshibashiWald}. 
This does not, of course, mean that the
observed acceleration can currently be said to be due to backreaction: the
situations we considered here are very much toy models (albeit 
ones that we expect to capture some of the properties of the real
Universe).  More study is required and, in particular, more realistic, 
non-perturbative models of the Universe are required, before any
definite conclusions can be drawn about the real Universe. One recent attempt 
at constructing such models appears to show some evidence of the effects
we describe here \cite{BolejkoFerreira}, but it is still neither conclusive nor 
fully realistic.

As a corollary of our study, a possible source of observational evidence for 
the hypothesis that
the apparent acceleration of the Universe is due to inhomogeneity
presents itself:  If the parameters of FLRW models inferred from local
observations are significantly different from those inferred from
observations made over large distances, then this would seem to imply
that the FLRW model that we use to model local spacetime is
different to the FLRW model that best describes the evolution of the
Universe on large scales.  Any such difference would signal a
significant departure from the predictions of the concordance $\Lambda$CDM model of the
Universe, and would therefore cast considerable doubt on the detection
of $\Lambda \neq 0$.  Of course, inferring cosmological parameters
from observations made on small scales is a considerable
challenge.   Sample variance due to the
presence of local structures, and the peculiar velocities they induce,
would have to be very well-understood.  Nevertheless, the supposed detection of
the `Hubble Bubble' \cite{Sinclair} suggests that it may not be entirely impossible.

The link we have found between observations and the spatial average can be 
explained by considering that, for a large enough collection of null rays, the 
typical conditions experienced by a ray at a given time, $t$, are likely to 
correspond to the average of
local conditions on a hypersurface of $t=$constant.  These averages are
exactly what Buchert's approach is constructed to estimate. As
long as  spacetime is statistically homogeneous and isotropic above
some scale, the result then follows (assuming observers and sources are 
distributed in a volume-weighted way). These issues have been
considered in detail in \cite{Rasanen3}.

In the case of the LTB model, in Section \ref{sect-results-ltb} we found that 
the Buchert average was more closely related to the local measures than to the 
typical distance-redshift relation of an observer, apparently in contradiction 
with our previous results. This model, however, is not statistically homogeneous or 
isotropic, and shows apparent (observational) acceleration only for a limited 
range of averaging domain sizes. Furthermore, the distance-redshift relation is 
only poorly represented by its monopole alone, since the dipole of the relation would 
certainly be important too. Therefore, our conclusions are that in this case 
the Buchert average is not enough to characterise the `typical' properties of 
the spacetime, and the mean of the monopole of the distance-redshift relation 
is also not enough to characterise what a typical observer should expect to see.
This leads one to question what it really means for an LTB model to exhibit 
`average acceleration' at all.

{\bf Acknowledgements:} We acknowledge the STFC and the BIPAC for
support, and A. Coley, C. Clarkson, P. G. Ferreira, S. R\"as\"anen, and 
O. Umeh for helpful discussions. We are particularly grateful to K. Bolejko for 
a thorough reading of the manuscript, and to the 
General Relativity and Cosmology group at Dalhousie University for their 
hospitality while part of this work was carried out. The computer code used 
in this paper is available online at \url{www.physics.ox.ac.uk/users/bullp/}.

%__________________________________________________

%________________________________________________

\end{document}